\documentclass[twocolumn]{aastex62}
\usepackage{amsmath}

\newcommand{\dst}{\mathrm{d}}
\newcommand{\sigmad}{\Sigma_{\dst}}
\newcommand{\sigmadup}{\Sigma_{\dst,0}}
\newcommand{\tstop}{t_{\mathrm{stop}}}
\newcommand{\taus}{\tau_{\mathrm{s}}}
\newcommand{\cs}{c_{\mathrm{s}}}
\newcommand{\cd}{c_{\mathrm{d}}}
\newcommand{\hd}{H_{\mathrm{d}}}
\def\msun{M_{\odot}}

\received{February 8, 2019}
\accepted{May 29, 2019 }
\submitjournal{ApJ}

\shorttitle{Turbulent diffusion and secular instabilities}
\shortauthors{Tominaga et al.}

\begin{document}

\title{REVISED DESCRIPTION OF DUST DIFFUSION AND A NEW INSTABILITY CREATING MULTIPLE RINGS IN PROTOPLANETARY DISKS}

\correspondingauthor{Ryosuke T. Tominaga}
\email{tominaga.ryosuke@a.mbox.nagoya-u.ac.jp}

\author[0000-0002-8596-3505]{Ryosuke T. Tominaga}
\affil{Department of Physics, Nagoya University, Nagoya, Aichi 464-8692, Japan}

\author{Sanemichi Z. Takahashi}
\affiliation{Department of Applied Physics, Kogakuin University, Hachioji, Tokyo, 192-0015, Japan}
\affiliation{National Astronomical Observatory of Japan, Osawa, Mitaka, Tokyo 181-8588, Japan}

\author{Shu-ichiro Inutsuka}
\affiliation{Department of Physics, Nagoya University, Nagoya, Aichi 464-8692, Japan}



\begin{abstract}
Various instabilities have been proposed as a promising mechanism to accumulate dust. Moreover, some of them are expected to lead to the multiple-ring structure formation and the planetesimal formation in protoplanetary disks. In a turbulent gaseous disk, the growth of the instabilities and the dust accumulation are quenched by turbulent diffusion of dust grains. The diffusion process has been often modeled by a diffusion term in the continuity equation for the dust density. The dust diffusion model, however, does not guarantee the angular momentum conservation in a disk. In this study, we first formulate equations that describe the dust diffusion and also conserve the total angular momentum of a disk. Second, we perform the linear perturbation analysis on the secular gravitational instability (GI) using the equations. The results show that the secular GI is a monotonically growing mode, contrary to the result of previous analyses that found it overstable. We find that the overstability is caused by the non-conservation of the angular momentum. Third, we find a new axisymmetric instability due to the combination of the dust-gas friction and the turbulent gas viscosity, which we refer to as two-component viscous gravitational instability (TVGI). The most unstable wavelength of TVGI is comparable to or smaller than the gas scale height. TVGI accumulates dust grains efficiently, which indicates that TVGI is a promising mechanism for the formation of multiple-ring-like structures and planetesimals. Finally, we examine the validity of the ring formation via the secular GI and TVGI in the HL Tau disk and find both instabilities can create multiple rings whose width is about 10 au at orbital radii larger than 50 au.
\end{abstract}



\section{Introduction}
Planets are thought to form through a process in which dust grains accumulate and undergo collisional growth in protoplanetary disks. Planetesimals are 10 km sized objects that form during the growth from dust grains to planets. Understanding the planetesimal formation is important in order to reveal the planet formation. 
Various instabilities are proposed as a possible mechanism to form planetesimals. The streaming instability is one example \citep[e.g.,][]{Youdin2005,Youdin2007,Johansen2007}. The streaming instability is driven by friction between gas and drifting dust, resulting in the dust-clump formation. Planetesimals are expected to form if the resulting dust clumps self-gravitationally collapse \citep[e.g.,][]{Johansen2007nature}. Secular gravitational instability (GI) is another possible mechanism of the planetesimal formation \citep[e.g.,][]{Ward2000,Youdin2011,Michikoshi2012,Takahashi2014}. The secular GI grows as a result of the decrease of the Coriolis force on the dust due to the dust-gas friction in a self-gravitationally stable disk. Since the growth of the secular GI also results in accumulation of dust grains, this instability has been proposed as the formation mechanism of planetesimals in outer disks and debris disks \citep[][]{Takahashi2014,Tominaga2018}. Moreover, the secular GI is one of the possible mechanisms to create multiple rings that recent observations with Atacama Large Millimeter/submillimeter Array (ALMA) have found in some protoplanetary disks \citep[e.g.,][]{ALMA-Partnership2015,Andrews2016,Tsukagoshi2016,Isella2016,Fedele2017,Fedele2018}. Based on the linear analysis, \citet{Takahashi2016} showed that the multiple rings in the HL Tau disk can form via the growth of the secular GI.

Protoplanetary disks are thought to be turbulent, resulting in the angular momentum transport in gas disks. Various hydrodynamic and magnetohydrodynamic instabilities have been proposed as mechanisms for driving the turbulence. These include the convective instabilitiy \citep[e.g.,][]{Lin1980,Klahr2014,Lyra2014}, the vertical shear instability \citep[e.g.,][]{Urpin1998,Urpin2003,Nelson2013} and the magnetorotational instability \citep[e.g.,][]{Balbus1991,Balbus1998}. It is known that dust grains diffuse in such a turbulent gas disk, and the diffusion coefficient is evaluated based on numerical simulations and analyses using the Langevin equations \citep[e.g.,][]{Carballido2005,Johansen2005,YL2007}. The dust diffusion is the most crucial process against the accumulation of dust grains. In previous work on the secular GI, the dust diffusion is modeled by introducing a diffusion term in the continuity equation for the dust, which is used for the linear analysis \citep[e.g.,][]{Youdin2011,Shariff2011,Takahashi2014,Shadmehri2016,Latter2017}. Although this modeling has been widely used \citep[e.g.,][]{Cuzzi1993,Goodman2000,DullemondPenzlin2018}, the modeling has a problem that the total angular momentum of the dusty gas disk is not conserved even when we consider the back-reaction from the dust to the gas and solve equations for both compnents. As described in Section \ref{sec:sec2}, the non-conservation of the angular momentum affects mainly on the motion of the dust. Hence, we need to resolve this problem to precisely discuss the accumulation process of dust grains. \citet{Goodman2000} also pointed out this problem. They artificially modified the momentum equations for the dust and avoided the problem arising from the diffusion term. This modification, however, is not validated. In this work, we first formulate phenomenological equations that guarantee the conservation of the angular momentum. Our formulation is based on the Reynold-averaging, which was also used in \citet{Cuzzi1993} and \citet{Shariff2011}. Next, we perform the linear analysis and investigate the stability of dusty-gas disks by using the formulated equations,.

This paper is organized as follows. In Section \ref{sec:sec2}, we show the problem mentioned above and describe how to resolve this problem. Basic equations are presented in Section \ref{sec:basiceq}. We show results of the linear analysis for cases with and without the turbulent viscosity of gas. We find a new instability driven by the combination of the friction and the turbulent viscosity besides the secular GI. In Section \ref{sec:discussion}, we discuss effect of the disk thickness and the multiple-ring formation via those instabilities.

\section{Angular momentum transport due to turbulent diffusion}\label{sec:sec2}

First of all, we show that equations used in the previous work \citep[e.g.,][]{Takahashi2014,Shadmehri2016,Latter2017,DullemondPenzlin2018} do not conserve the total angular momentum of a dusty gas disk. We assume that a disk is infinitesimally thin and axisymmetric and solve the evolutionary equations of the vertically integrated disk. We adopt a cylindrical coordinate system $(r,\phi)$. The continuity equation and the azimuthal equation of motion for the dust are as follows:
\begin{equation}
\frac{\partial\sigmad}{\partial t}+\frac{1}{r}\frac{\partial\left(r\sigmad v_r\right)}{\partial r}=\frac{1}{r}\frac{\partial}{\partial r}\left(rD\frac{\partial\sigmad}{\partial r}\right),\label{eq:eocdust}
\end{equation}
\begin{equation}
\sigmad\left[\frac{\partial v_{\phi}}{\partial t}+v_r\frac{\partial v_{\phi}}{\partial r}\right]=-\sigmad\frac{v_{\phi}v_r}{r}-\sigmad\frac{v_{\phi}-u_{\phi}}{\tstop},
\end{equation}
where $\sigmad$ is the surface density of the dust, $v_r$ and $v_{\phi}$ are the radial and azimuthal velocities of the dust, respectively. The diffusion coefficient is denoted by $D$, and the azimuthal velocity of the gas is $u_{\phi}$. The stopping time of a dust grain is $\tstop$. The term on the right hand side of Equation (\ref{eq:eocdust}) is often introduced to model the dust diffusion in a turbulent disk \citep[e.g.,][]{Youdin2011}. From these equations, we obtain an equation governing the evolution of the angular momentum of the dust $\sigmad j_{\dst}\equiv\sigmad rv_{\phi}$:
\begin{align}
\frac{\partial\left(\sigmad j_{\dst}\right)}{\partial t}&+\frac{1}{r}\frac{\partial}{\partial r}\left(rv_r\sigmad j_{\dst}\right)\notag\\
&=-r\sigmad\frac{v_{\phi}-u_{\phi}}{\tstop}+j_{\dst}\frac{1}{r}\frac{\partial}{\partial r}\left(rD\frac{\partial\sigmad}{\partial r}\right).\label{eq:angmondust_previous}
\end{align}
The first term on the right hand side stands for the angular momentum transport due to the friction between the dust and the gas. If we take into account the back-reaction from the dust to the gas and the gas motion, this friction term does not violate the conservation of the total angular momentum of a dusty gas disk, as shown in the following. The continuity equation and the azimuthal equation of motion for gas are the following:
\begin{equation}
\frac{\partial \Sigma}{\partial t}+\frac{1}{r}\frac{\partial\left(r\Sigma u_r\right)}{\partial r}=0,\label{eq:eocgas}
\end{equation}
\begin{equation}
\Sigma\left[\frac{\partial u_{\phi}}{\partial t}+u_r\frac{\partial u_{\phi}}{\partial r}\right]=-\Sigma\frac{u_{\phi}u_r}{r}+\sigmad\frac{v_{\phi}-u_{\phi}}{\tstop},
\end{equation}
where $\Sigma$ and $u_r$ are the surface density and the radial velocity of the gas, respectively. We obtain an equation describing the time evolution of the angular momentum of the gas $\Sigma j_{\mathrm{g}}\equiv\Sigma ru_{\phi}$ as follows: 
\begin{equation}
\frac{\partial\left(\Sigma j_{\mathrm{g}}\right)}{\partial t}+\frac{1}{r}\frac{\partial}{\partial r}\left(ru_r\Sigma j_{\mathrm{g}}\right)=r\sigmad\frac{v_{\phi}-u_{\phi}}{\tstop}.\label{eq:angmongas_previous}
\end{equation}
Equations (\ref{eq:angmondust_previous}) and (\ref{eq:angmongas_previous}) give the following equation:
\begin{align}
\frac{\partial}{\partial t}\left(\Sigma j_{\mathrm{g}}+\sigmad j_{\dst}\right)&+\frac{1}{r}\frac{\partial}{\partial r}\left(ru_r\Sigma j_{\mathrm{g}}+rv_r\sigmad j_{\dst}\right)\notag\\
&=j_{\dst}\frac{1}{r}\frac{\partial}{\partial r}\left(rD\frac{\partial\sigmad}{\partial r}\right).\label{eq:angmon_change}
\end{align}
The term on the right hand side cannot be written in a form of the divergence of the angular momentum flux because it is proportional to the specific angular momentum. Thus, the volume integral of this term is not zero in general, meaning that the total angular momentum of the disk is not conserved. This is due to the treatment of the dust diffusion. The diffusion term introduced in the continuity equation for the dust (Equation (\ref{eq:eocdust})) unphysically changes the angular momentum of the dust disk and prevents the total angular momentum from being conserved. The gas motion is also affected by the unphysical change of the angular momentum of the dust because it is partly transported to the gas through the dust-gas friction. The effect on the gas motion is, however, smaller than that on the dust motion by a factor of the dust-to-gas mass ratio because the rate of the angular momentum transport due to the friction is proportional to the surface density of the dust (Equation (\ref{eq:angmongas_previous})).
This implies that the non-conservation of the angular momentum affects mainly on the motion of the dust. 

In order to discuss how the non-conservation of the angular momentum of a dusty gas disk affects the motion of the dust, we rearrange Equation (\ref{eq:angmondust_previous}) as follows:
\begin{align}
\frac{\partial\left(\sigmad j_{\dst}\right)}{\partial t}&+\frac{1}{r}\frac{\partial}{\partial r}\left[r\left(v_r-\frac{D}{\sigmad}\frac{\partial \sigmad}{\partial r}\right)\sigmad j_{\dst}\right]\notag\\
&=-r\sigmad\frac{v_{\phi}-u_{\phi}}{\tstop}-D\frac{\partial \sigmad}{\partial r}\frac{\partial j_{\dst}}{\partial r}.\label{eq:angmondust_prev_rearranged}
\end{align}
We do not consider the effect of the second term on the left hand side  since this term becomes the surface term and vanishes when it is integrated over all space. The second term on the right hand side changes the total angular momentum of the disk. If we consider that the disk rotates with the Keplerian velocity, that term represents negative (positive) torque exerted on the dust when the surface density gradient of the dust is positive (negative). In the dust accumulating region, for example, the dust in the inner part ($\partial\sigmad/\partial r>0$) loses the angular momentum and goes inward, and vice versa. This unphysical torque prevents the accumulation of the dust. In other words, the previous studies underestimated the degree of dust accumulation. Hence, in order to discuss the accumulation process precisely, we need to re-formulate equations that conserve the angular momentum.

If the dust grains are small, and the stopping time is short compared to the Keplerian period, the diffusion is mainly driven by radial kicks from the turbulent gas \citep{YL2007}, meaning that the specific angular momentum of the dust grains does not change along the diffusion flow. Equation (\ref{eq:eocdust}) is rearranged to yield
\begin{equation}
\frac{\partial \sigmad}{\partial t}+\frac{1}{r}\frac{\partial}{\partial r}\left[r\left(v_r-\frac{D}{\sigmad}\frac{\partial \sigmad}{\partial r}\right)\sigmad\right]=0,
\end{equation}
which means that the advection velocity caused by the diffusion is written by $-D\sigmad^{-1}\partial\sigmad/\partial r$. Then, we consider a case that the time evolution of the specific angular momentum is governed by the following equation:
\begin{equation}
\sigmad\left[\frac{\partial j_{\dst}}{\partial t}+\left(v_r-\frac{D}{\sigmad}\frac{\partial\sigmad}{\partial r}\right)\frac{\partial j_{\dst}}{\partial r}\right]=-r\sigmad\frac{v_{\phi}-u_{\phi}}{\tstop}.\label{eq:speang_rphi}
\end{equation}
(see Appendix \ref{ap:formulation} for the detailed derivation). From this equation, we obtain the following equation for the angular momentum of the dust:
\begin{align}
\frac{\partial\left(\sigmad j_{\dst}\right)}{\partial t}&+\frac{1}{r}\frac{\partial}{\partial r}\left[r\left(v_r-\frac{D}{\sigmad}\frac{\partial\sigmad}{\partial r}\right)\sigmad j_{\dst}\right]\notag\\
&=-r\sigmad\frac{v_{\phi}-u_{\phi}}{\tstop}.\label{eq:angmonchange_sec2}
\end{align}
This equation shows that the angular momentum of the dust disk changes only through the friction. From Equations (\ref{eq:angmongas_previous}) and (\ref{eq:angmonchange_sec2}), we can show that the total angular momentum of the dusty-gas disk is conserved. In this way, we see that the equations conserve the total angular momentum if the advection velocity by the diffusion is considered in the equation for the specific angular momentum. Such an advection term is naturally derived by the Reynolds averaging of hydrodynamic equations for the dust. We also formulate a radial momentum equation consistent with Equations (\ref{eq:eocdust}) and (\ref{eq:speang_rphi}) based on the Reynolds averaging in Appendix \ref{ap:formulation}. We summarize a set of basic equations including the newly formulated dust equations in the next section.

\section{Basic equations}\label{sec:basiceq}
In this work, we perform the linear stability analysis of an infinitesimally thin and axisymmetric disk with taking into account the conservation of the angular momentum in the disk as discussed in the previous section. We summarize the basic equations for gas and dust in this section. We use the following equations for the gas and the Poisson equation, which were also used in \citet{Takahashi2014, Takahashi2016}: 
\begin{equation}
\frac{\partial \Sigma}{\partial t}+\frac{1}{r}\frac{\partial\left(r\Sigma u_r\right)}{\partial r}=0,\tag{\ref{eq:eocgas}}
\end{equation}
\begin{align}
\Sigma\left(\frac{\partial u_i}{\partial t}+u_j\frac{\partial u_i}{\partial x_j}\right)&=-\cs^2\frac{\partial \Sigma}{\partial x_i}-\Sigma\frac{\partial}{\partial x_i}\left(\Phi-\frac{GM_{\ast}}{r}\right)\notag\\
&+\frac{\partial}{\partial x_j}\left[\Sigma\nu\left(\frac{\partial u_i}{\partial x_j}+\frac{\partial u_j}{\partial x_i}-\frac{2}{3}\delta_{ij}\frac{\partial u_k}{\partial x_k}\right)\right]\notag\\
&+\sigmad\frac{v_i-u_i}{\tstop},\label{eq:eomgas}
\end{align}
\begin{equation}
\nabla^2\Phi=4\pi G\left(\Sigma+\sigmad\right)\delta(z),\label{eq:poisson}
\end{equation}
where $u_i,v_i$ are the $i$-th component of the velocities of the gas and the dust, $\cs$ is the sound speed, $\Phi$, $G$ and $M_{\ast}$ are the gravitational potential of the disk, the gravitational constant and the mass of the central star, respectively. We denote the coefficient of the turbulent viscosity by $\nu$ that is measured by the dimensionless parameter $\alpha\equiv\nu\Omega\cs^{-2}$ \citep{Shakura1973}, where $\Omega$ is the angular velocity of the gas disk. 

Next, we summarize equations for the dust. We derived the following equations for the dust, which include Equation (\ref{eq:eocdust}), based on the Reynolds averaging:
\begin{equation}
\frac{\partial\sigmad}{\partial t}+\frac{1}{r}\frac{\partial\left(r\sigmad v_r\right)}{\partial r}=\frac{1}{r}\frac{\partial}{\partial r}\left(rD\frac{\partial\sigmad}{\partial r}\right),\tag{\ref{eq:eocdust}}
\end{equation}
\begin{align}
\sigmad\biggl[\frac{\partial v_r}{\partial t}+&\left(v_r-\frac{D}{\sigmad}\frac{\partial\sigmad}{\partial r}\right)\frac{\partial v_r}{\partial r}\biggr]\notag\\
=&\sigmad\frac{v_{\phi}^2}{r}-\cd^2\frac{\partial\sigmad}{\partial r}-\sigmad\frac{\partial}{\partial r}\left(\Phi-\frac{GM_{\ast}}{r}\right)\notag\\
&-\sigmad\frac{v_r-u_r}{\tstop}+\frac{1}{r}\frac{\partial}{\partial r}\left(rv_rD\frac{\partial\sigmad}{\partial r}\right),\label{eq:eomrdust}
\end{align}
\begin{align}
\sigmad\biggl[\frac{\partial v_{\phi}}{\partial t}&+\left(v_r-\frac{D}{\sigmad}\frac{\partial\sigmad}{\partial r}\right)\frac{\partial v_{\phi}}{\partial r}\biggr]\notag\\
&=-\sigmad\frac{v_{\phi}}{r}\left(v_r-\frac{D}{\sigmad}\frac{\partial\sigmad}{\partial r}\right)-\sigmad\frac{v_{\phi}-u_{\phi}}{\tstop},\label{eq:eomphidust}
\end{align}
where $\cd$ is the velocity dispersion of the dust. The advection velocity $-D\sigmad^{-1}\partial\sigmad/\partial r$ that appears on the left hand side of Equation (\ref{eq:eomrdust}) and on the both sides of Equation (\ref{eq:eomphidust}) is not included in the previous studies. The last term on the right hand side of Equation (\ref{eq:eomrdust}) is also not included. See Appendix \ref{ap:formulation} for detailed derivation. From Equations (\ref{eq:eocdust}), (\ref{eq:eocgas}), (\ref{eq:eomgas}) and (\ref{eq:eomphidust}), we can derive the evolutionary equation for the total angular momentum:
\begin{align}
\frac{\partial}{\partial t}\left(\Sigma j_{\mathrm{g}}+\sigmad j_{\dst}\right)&+\frac{1}{r}\frac{\partial}{\partial r}\left[ru_r\Sigma j_{\mathrm{g}}+r\left(v_r-\frac{D}{\sigmad}\frac{\partial\sigmad}{\partial r}\right)\sigmad j_{\dst}\right]\notag\\
&=\frac{1}{r}\frac{\partial}{\partial r}\left(r^3\Sigma\nu\frac{\partial \Omega}{\partial r}\right),\label{eq:new_angmon_change_w_vis}
\end{align}
The term on the right hand side shows the angular momentum transport due to the viscosity. Thus, our basic equations conserve the total angular momentum even if we take into account the dust diffusion.

We perform the linear analysis in local Cartesian coordinates $(x,y)=(r-R_0,R_0(\phi-\Omega t))$ corotating with the angular velocity of the disk $\Omega$ at the radius $R_0$. As an unperturbed state, we assume that the surface densities of the gas and the dust are uniform,  $u_{x,0}=v_{x,0}=0$, and $u_{y,0}=v_{y,0}=-(3/2)\Omega x$. Considering axisymmetric perturbations $\delta\Sigma,\delta\sigmad,\delta u_x,\delta u_y, \delta v_x, \delta v_y,\delta \Phi$ proportional to $\exp[nt+ikx]$, we linearize Equations (\ref{eq:eocdust}), (\ref{eq:eocgas}), (\ref{eq:eomgas})--(\ref{eq:eomphidust}). The linearized equations are given as follows:
\begin{equation}
n\delta\Sigma+ik\Sigma_0\delta u_x=0,\label{eq:lin-eocgas}
\end{equation}
\begin{equation}
n\delta u_x=2\Omega\delta u_y-\frac{\cs^2}{\Sigma_0}ik\delta\Sigma-ik\delta\Phi-\frac{4}{3}\nu k^2\delta u_x+\epsilon\frac{\delta v_x-\delta u_x}{\tstop},\label{eq:lin-eomxgas}
\end{equation}
\begin{equation}
n\delta u_y=-\frac{\Omega}{2}\delta u_x-\nu k^2\delta u_y-ik\frac{3\nu\Omega}{2\Sigma_0}\delta\Sigma+\epsilon\frac{\delta v_y-\delta u_y}{\tstop},\label{eq:lin-eomygas}
\end{equation}
\begin{equation}
n\delta\sigmad+ik\sigmadup\delta v_x=-Dk^2\delta\sigmad,\label{eq:lin-eocdust}
\end{equation}
\begin{equation}
n\delta v_x=2\Omega\delta v_y-\frac{\cd^2}{\sigmadup}ik\delta\sigmad-ik\delta\Phi-\frac{\delta v_x-\delta u_x}{\tstop},\label{eq:lin-eomxdust}
\end{equation}
\begin{equation}
n\delta v_y=-\frac{\Omega}{2}\left(\delta v_x-\frac{ikD}{\sigmadup}\delta\sigmad\right)-\frac{\delta v_y-\delta u_y}{\tstop},\label{eq:lin-eomydust}
\end{equation}
\begin{equation}
\delta\Phi=-\frac{2\pi G}{k}\left(\delta\Sigma+\delta\sigmad\right),\label{eq:lin-poisson}
\end{equation}
where $n$ is the growth rate, $k$ is the wavenumber, and $\epsilon\equiv\sigmadup/\Sigma_0$ is the dust-to-gas mass ratio. We do not consider the large-scale pressure gradient in the background state or the vertical motion of the gas and the dust in this work, which precludes the streaming instability. This setup is suitable to study growing modes different from the streaming instability.

\begin{figure*}
	\begin{center}
		\includegraphics[width=1.5\columnwidth]{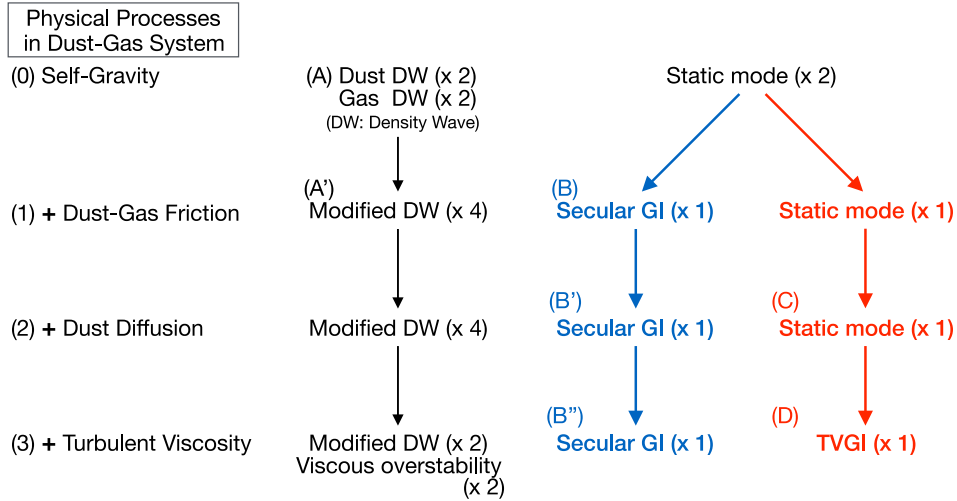}
	\end{center}
    		\caption{A schematic diagram to classify the modes. There are six modes in our analysis. The modes on the top line are those we obtain when we neglect the dust-gas friction, the dust diffusion, and the turbulent gas viscosity (see Appendix \ref{ap:staticmode}). (Step 1, the second line) The dust-gas friction couples the dust and gas density waves, resulting in modified density waves (DWs, the mode A'). If the self-gravity is strong enough, the modified density waves become unstable, which are referred to as the classical GI. One of the static mode becomes the secular GI by the friction (the mode B, see Section \ref{subsec:wovis}). (Step 2, the third line) The dust diffusion does not qualitatively change properties of those six modes although the growth rate for each mode is changed. (Step 3, the bottom line) The turbulent viscosity of the gas destabilizes the static mode remaining in Step 2 (the mode D). This destabilized static mode is referred to as Two-component Viscous Gravitational Instability or TVGI (see Section \ref{subsec:wvis}). Two of the modified DWs modes become the viscous overstable mode.   
			}
   		 \label{fig:modeclass}
\end{figure*}
\section{Results}\label{sec:linearanalysis}
In this section, we present the results of the linear analyses with and without the turbulent viscosity and compare our results with those of the previous work. There are six modes in our analysis because we solve the equations for both gas and dust as described above. In the most simplified limit for the gas and the dust where we do not consider any of the dust-gas friction, the dust diffusion, or the turbulent gas viscosity, these are two density waves for each component and two static modes, and hence six modes in total (see Appendix \ref{ap:staticmode}). A static mode is a steady solution of the perturbation equations. Figure \ref{fig:modeclass} shows how the six modes change by adding three physical processes step by step and which mode becomes unstable. As shown in Figure \ref{fig:modeclass}, the two static modes become unstable, which is explained in more detail in the following subsections. 

\begin{figure*}
	\begin{tabular}{c}
		\begin{minipage}{0.5\hsize}
			\begin{center}
				\includegraphics[width=0.9\columnwidth]{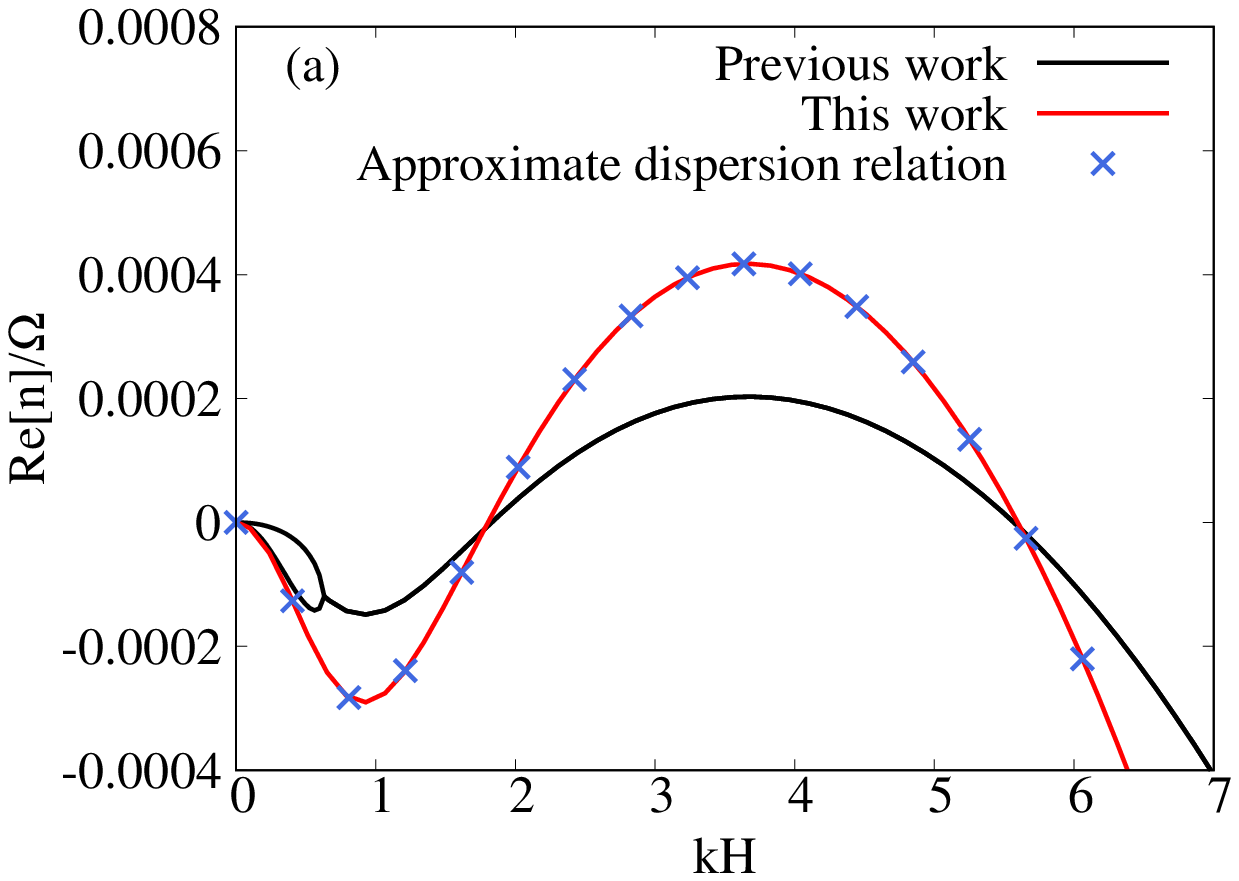}
			\end{center}
		\end{minipage}
		\begin{minipage}{0.5\hsize}
			\begin{center}
				\includegraphics[width=0.9\columnwidth]{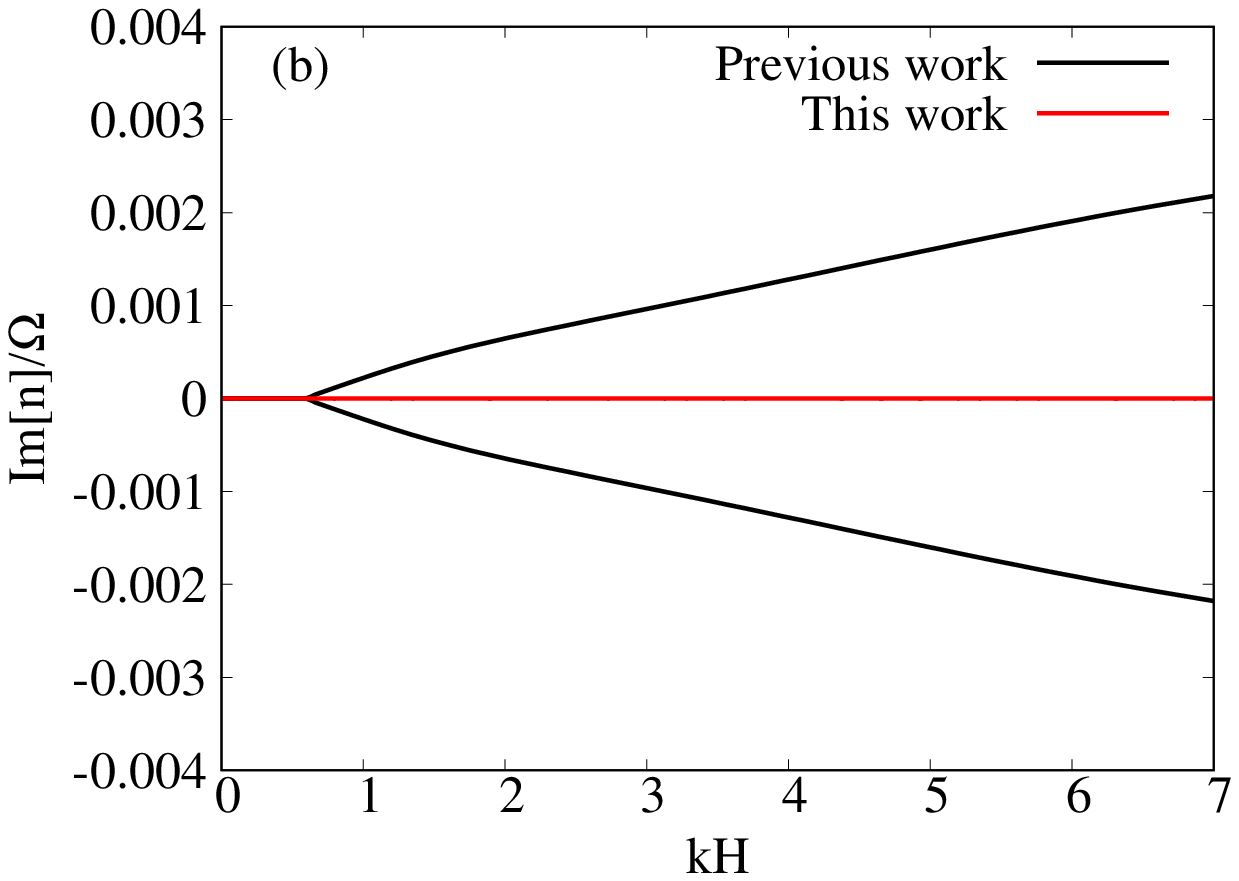}
			\end{center}
		\end{minipage}
	\end{tabular}
\caption{Dispersion relation of the secular GI for $D=10^{-4}\cs^2\Omega^{-1},\cd=0,\epsilon=0.1,\taus=0.01,Q=3$. The horizontal axis of both panels is the wavenumber normalized by the gas scale height $H\equiv\cs/\Omega$. The vertical axis of the left panel is the real part of $n$ normalized by the angular velocity $\Omega$, and that of the right panel is the imaginary part. The black line is the dispersion relation obtained in the previous work \citep{Takahashi2014,Latter2017}, while the red line shows the result of our analysis. The blue cross mark is the dispersion relation based on the terminal velocity approximation (Equation (\ref{eq:NEWapdis})). The imaginary part $\mathrm{Im}[n]$ is zero for the secular GI mode obtained in our analysis in contrast to the previous work.}
 \label{fig:diff00001_cd0_ep01_ts001_Q3}
\end{figure*}
\subsection{Without Turbulent Viscosity}\label{subsec:wovis}
First, we show results obtained when we neglect the viscosity term in Equation (\ref{eq:eomgas}), which corresponds to Step (2) in Figure \ref{fig:modeclass}. We note that we relate $D$ and $\cd$ to the dimensionless turbulent strength $\alpha$ as shown below (Equations (\ref{YLdiff}) and (\ref{YLcd})) although $\nu$ does not appear in this section. This is because it is easier to compare the results with those in Section \ref{subsec:wvis}.

We find one static mode ($n=0$) and one mode that can be unstable (the modes B' and C in Figure \ref{fig:modeclass}). The static mode is the perturbed state where the dust has the same azimuthal velocity with the gas, and the radial force balance holds. The latter mode corresponds to the secular GI. By using the terminal velocity approximation ($\tstop\ll n^{-1}$) \footnote{The terminal velocity approximation sometimes refers to $v_{i}=u_i+\tstop\Sigma^{-1}\partial \left(\cs^2\Sigma\right)/\partial x_i$. This expression is valid only when the Coriolis force and the gas viscosity can be neglected. ``The terminal velocity approximation" adopted here is more general one, which refers to neglecting the time derivative of the relative velocity by assuming it is small enough compared to the friction term. } and assuming $\tstop\ll\Omega^{-1}\ll n^{-1}$, we obtain the following approximate dispersion relation of the secular GI:
\begin{align}
&A_1n+A_0=0,\label{eq:NEWapdis}\\
&A_1\equiv\left\{\Omega^2+\left(\frac{1+\epsilon}{\tstop} \right)^2\right\}\omega_{\mathrm{gd}}^2+\frac{\epsilon D k^2}{\tstop}\cs^2k^2,\\
&A_0\equiv\frac{\omega_{\mathrm{g}}^2}{1+\epsilon}\left\{\left(\frac{1+\epsilon}{\tstop}\right)^2Dk^2+\frac{1+\epsilon}{\tstop}\cd^2k^2\right\}+\omega_{\dst}^2\frac{\epsilon c_{\mathrm{s}}^2k^2}{\tstop},
\end{align}
where
\begin{align}
\omega_{\mathrm{gd}}^2&\equiv \Omega^2+\frac{\cs^2+\epsilon\cd^2}{1+\epsilon}k^2-2\pi G\left(1+\epsilon\right)\Sigma_0k,\\
\omega_{\mathrm{g}}^2&\equiv\Omega^2+\cs^2k^2-2\pi G\left(1+\epsilon\right)\Sigma_0k,\\
\omega_{\dst}^2&\equiv\Omega^2+\cd^2k^2-2\pi G\left(1+\epsilon\right)\Sigma_0k.
\end{align}

Figure \ref{fig:diff00001_cd0_ep01_ts001_Q3} shows the dispersion relation of the secular GI obtained in this work and that of the previous work for $D=10^{-4}\cs^2\Omega^{-1},\cd=0,\epsilon=0.1,\taus\equiv\tstop\Omega=0.01$ and $Q\equiv\cs\Omega/\pi G\Sigma_0=3$. As mentioned in \citet{Takahashi2014}, the secular GI mode obtained in this work is also stabilized for the long wavelength perturbation by the Coriolis force exerted on the dust. In contrast, the short wavelength perturbation is stabilized by the turbulent diffusion. For wavelengths where the secular GI mode is unstable, the gas pressure gradient force dominates the Coriolis force acting on the gas as a stabilizing force, which forms an azimuthal zonal flow \citep[see,][]{Latter2017}, and the dust accumulates by the self-gravity of itself. The growth rate of the secular GI obtained in this analysis is several times larger than that obtained in the previous work. In addition, we find the secular GI is a monotonically growing mode, while the secular GI found in the previous work is overstable depending on parameters\footnote{The approximated dispersion relation, Equation (\ref{eq:NEWapdis}), is linear for $n$, although that derived in the previous work \cite[Equation (13) in ][]{Takahashi2014} is a quadratic equation. This is why the overstable mode does not appear in our formulation.}. This shows that the overstable mode obtained in the previous work is due to the unphysical torque acting on the dust. We derive an approximate condition for the growth of the secular GI from Equation (\ref{eq:NEWapdis}). The condition is that at least two solutions for $n=0$ exist in the region $k>0$, which is equivalent to the condition that the equation $A_0/k^2=0$ has two distinct positive real solutions. We then obtain the following approximate condition for the instability:
\begin{equation}
\frac{Q^2\left(\tstop\cd^2+D\right)}{\left(1+\epsilon\right)\left[\tstop\left(\epsilon\cs^2+\cd^2\right)+D\left(1+\epsilon\right)\right]}<1.\label{eq:app_unstcond_NEWSGI}
\end{equation}
Equation (\ref{eq:app_unstcond_NEWSGI}) is equivalent to the condition derived in \citet{Latter2017} in the case of $D=0$ (see Equation (43) in their paper). If we assume $\cd=0$ and $(1+\epsilon)D\ll \epsilon\cs^2\tstop$, Equation (\ref{eq:app_unstcond_NEWSGI}) becomes
\begin{equation}
Q<\sqrt{\frac{\epsilon\left(1+\epsilon\right)\tstop\cs^2}{D}},\label{eq:app_cond_limit_case}
\end{equation}
which is equivalent to the condition obtained in the previous work. For instance, when $(1+\epsilon)D\ll \epsilon\cs^2\tstop$ is satisfied, Equation (50) in \citet[][]{Latter2017} is equivalent to the above condition (Equation (\ref{eq:app_cond_limit_case})). In this way, we find that our formulation does not change the condition for the secular GI if $D$ is small enough to satisfy $(1+\epsilon)D\ll \epsilon\cs^2\tstop$.
\begin{figure*}
	\begin{center}
		\includegraphics[width=1.5\columnwidth]{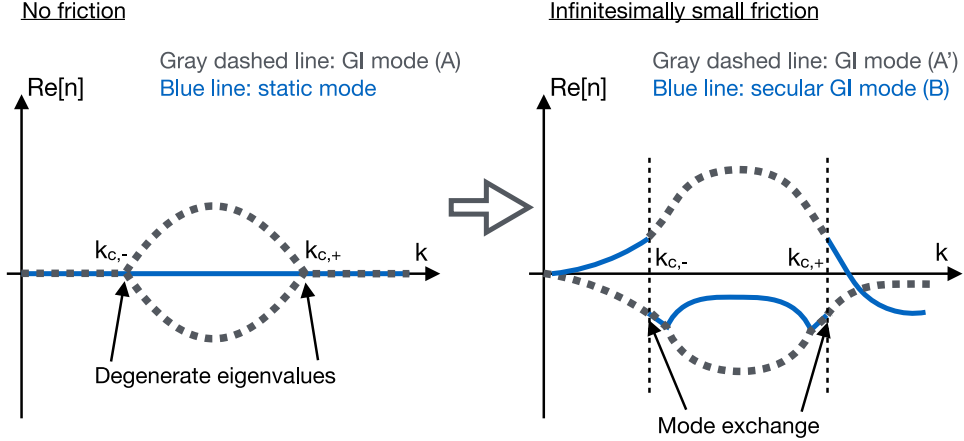}
	\end{center}
    		\caption{Schematic picture to show how the mode exchange occurs between the dust GI and the secular GI. On the left figure, we show the dispersion relations of the dust GI and the static mode for the case without the friction. The right figure shows the dispersion relations obtained with the friction force. The gray dashed line represents the growth rate of the dust GI mode, and the blue solid line is that of the static mode or the secular GI. The labels (A), (A') and (B) shown in the legends correspond to the labels shown in Figure \ref{fig:modeclass}. The mode exchange occurs at the wavelengths where the eigenvalue and eigenfunction degenerate for the case without the friction.} 
   		 \label{fig:mode_exchange-SGI-GI}
\end{figure*}

\begin{figure}
	\begin{center}
		\includegraphics[width=\columnwidth]{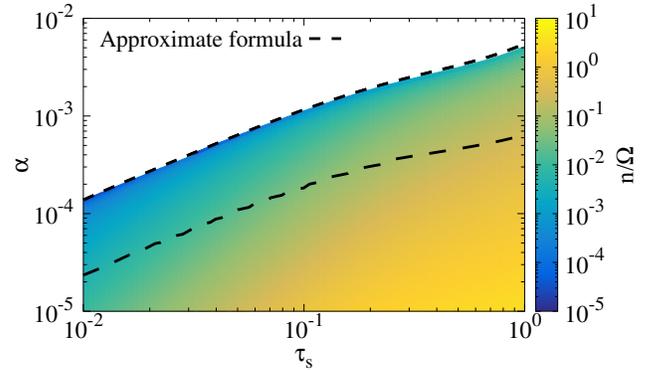}
	\end{center}
    		\caption{Maximum growth rate of the instabilities for $\epsilon=0.1$ and $Q=3$. The horizontal axis is the normalized stopping time $\taus$, and the vertical axis is the strength of turbulence $\alpha$. The color represents the maximum growth rate normalized by the angular velocity $\Omega$. We note that $\mathrm{Im}[n]$ is zero in the whole parameter space. The short-dashed line represents the approximate condition for the instability (Equation (\ref{eq:app_unstcond_NEWSGI})). The dust GI is the most unstable mode below the long-dashed line. In the colored region above the long-dashed line, the secular GI is the fastest growing mode. Both instabilities are stable in the white region.}
   		 \label{fig:Q3ep01-growthrate_wo_thickness_wo_viscosity}
\end{figure}

We find that the mode exchange between the secular GI mode and the classical GI mode, which is an unstable density wave (see Appendix \ref{ap:staticmode}), occurs at $k=k_{\mathrm{c},-},k_{\mathrm{c},+}$ where the growth rate of the classical GI mode becomes zero (Figure \ref{fig:mode_exchange-SGI-GI}). The mode exchange is reconnection of the curves in the $k$ -- $n$ plane of dispersion relations for two different modes. We note that the classical GI mode seen here is the dust GI mode mediated by the self-gravitationally stable gas disk. Hereafter, we simply refer to the classical GI mode as the dust GI mode in order to avoid confusion with the gas GI mode that does not appear since $Q$ is larger than unity in this paper. We can see the mode exchange even in the linear analysis only for the motion of the dust \citep[see Figure 10 in][]{Youdin2011}. As shown in Figure \ref{fig:mode_exchange-SGI-GI}, we designate the growing mode in the limited range of wavenumber where the dust GI remains unstable for $\tstop\to\infty$ (no friction between gas and dust) as ``the dust GI". On the other hand, we designate the growing mode in the disconnected curves in the regions of wavenumber where the dust GI is stable for $\tstop\to\infty$ as ``the secular GI".
Figure \ref{fig:Q3ep01-growthrate_wo_thickness_wo_viscosity} shows the maximum growth rate of the instabilities as a function of $\taus$ and $\alpha$ for $\epsilon=0.1,Q=3$, where we use the following equations to calculate the diffusion coefficient and the velocity dispersion \citep{YL2007}:
\begin{equation}
D=\frac{1+\taus+4\taus^2}{\left(1+\taus^2\right)^2}\alpha\frac{\cs^2}{\Omega},\label{YLdiff}
\end{equation}
\begin{equation}
\cd=\frac{\sqrt{1+2\taus^2+(5/4)\taus^3}}{1+\taus^2}\sqrt{\alpha}\cs.\label{YLcd}
\end{equation}
We also show the upper limit of $\alpha$ obtained from the approximate condition for the instability (Equation (\ref{eq:app_unstcond_NEWSGI})) in Figure \ref{fig:Q3ep01-growthrate_wo_thickness_wo_viscosity}. The exact upper limit of $\alpha$ is well represented by Equation (\ref{eq:app_unstcond_NEWSGI}). The secular GI is the fastest growing mode in the colored region above the long-dashed line, while the dust GI mode grows faster in the region below the long-dashed line. The most unstable wavenumber on the dashed line is $k_{\mathrm{c},-}$. The dust GI becomes the fastest growing mode for smaller $\alpha$ cases since the stabilizing effect of the diffusion and the velocity dispersion is smaller.

\subsection{With Turbulent Viscosity: Two-Component Viscous Gravitational Instability}\label{subsec:wvis}

Next, we discuss the linear stability in the case with the turbulent viscosity acting on the gas (Step 3 in Figure \ref{fig:modeclass}). In this case, we find a new instability that is different from the secular GI. We designate this new instability as two-component viscous gravitational instability (TVGI). TVGI did not appear in \citet{Takahashi2014}, in which the basic equations did not guarantee the conservation of the total angular momentum. The origin of TVGI is the static mode that appears in the case without the turbulent viscosity (the mode labeled (C) in Figure \ref{fig:modeclass}). The static mode is the steady solution where the gas and the dust have the same azimuthal velocity ($\delta u_y=\delta v_y$) as mentioned in Section \ref{subsec:wovis} and Appendix \ref{ap:staticmode}. In this steady state, the self-gravity balances mainly with the Coriolis force, that is,  $2\Omega\delta u_y -ik\delta\Phi=2\Omega\delta v_y -ik\delta\Phi\simeq 0$. This radial force balance is not realized once we consider the gas viscosity. The viscosity decreases $\delta u_y$, which results in the relative velocity in the azimuthal direction between the dust and the gas. The relative azimuthal velocity causes the decrease of $\delta v_y$ through the friction term. The decrease of both $\delta u_y$ and $\delta v_y$ prevents the radial force balance since the Coriolis force in the radial direction decreases. Thus, both dust and gas accumulate by the self-gravity, which is the physical interpretation of TVGI.
 The decrease of the Coriolis force by both viscosity and friction is important for the growth of TVGI. If we neglect the friction, this unstable mode becomes a static mode that satisfies the radial force balance for both components and
\begin{equation}
\delta u_x = 0, 
\end{equation}
\begin{equation}
-\nu k^2\delta u_y-ik\frac{3\nu\Omega}{2\Sigma_0}\delta\Sigma=0,
\end{equation}
\begin{equation}
\delta v_x-\frac{ikD}{\sigmadup}\delta\sigmad=0,
\end{equation}
(see Equations (\ref{eq:lin-eocgas}) -- (\ref{eq:lin-eomydust})). This indicates that TVGI is different from the so-called viscous instability, which grows in the one fluid system and does not need the friction \citep[e.g.,][]{Lynden-Bell1974,Schmit1995,Gammie1996,Lin2016}. 

\begin{figure}
	\begin{center}
		\includegraphics[width=\columnwidth]{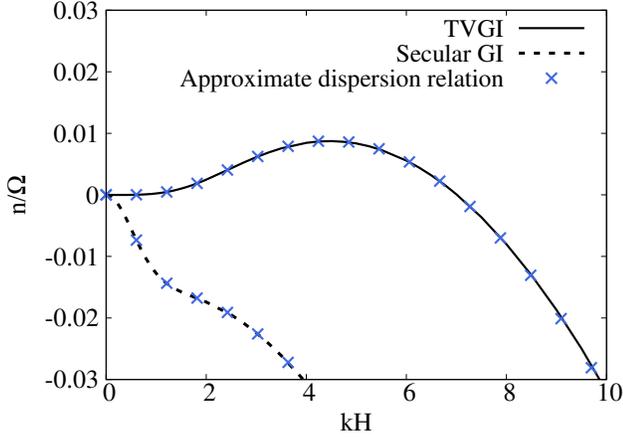}
	\end{center}
    		\caption{Dispersion relations of TVGI and the secular GI for $\alpha=10^{-3}, \epsilon=0.1, \taus=0.3$ and $Q=5$. The horizontal axis is the normalized wavenumber, and the vertical axis is the normalized growth rate $n$. The solid and dashed lines represent the dispersion relation of TVGI and the secular GI, respectively. The blue cross mark is the approximated dispersion relation (Equation (\ref{eq:TVGIapdis})). In this case, the secular GI is stable.}
   		 \label{fig:Q5ep01taus03alp1e-3_dispersion}
\end{figure}

By using the terminal velocity approximation and assuming $\tstop\ll\Omega^{-1}\ll n^{-1}$, the dispersion relation is reduced to the following quadratic equation:
\begin{equation}
B_2n^2+B_1n+B_0=0,\label{eq:TVGIapdis}
\end{equation}
\begin{align}
B_2\equiv&\left\{\Omega^2+\left(\frac{1+\epsilon}{\tstop}\right)^2\right\}\omega_{\mathrm{gd}}^2+\frac{\epsilon Dk^2}{\tstop}\cs^2k^2+\frac{\nu k^2}{\left(1+\epsilon\right)^2}\notag\\
&\times\biggl[\frac{\left(1+\epsilon\right)^2}{\tstop}\epsilon\omega_{\mathrm{gd}}^2+\epsilon Dk^2\left\{3\left(1+\epsilon\right)\Omega^2+\epsilon\cs^2k^2\right\}\notag\\
&+\frac{\left(1+\epsilon\right)^3}{\tstop^2}Dk^2+\frac{1+\epsilon}{\tstop}\left(\cd^2+\epsilon\cs^2\right)k^2\biggr]+\frac{4\nu k^2}{3}\frac{1+\epsilon}{\tstop}\notag\\
&\times\left(\frac{Dk^2}{\tstop}+\frac{\epsilon\Omega^2}{1+\epsilon}+\cd^2k^2-2\pi G\epsilon\Sigma_0k\right),\\
B_1\equiv&\frac{\omega_{\mathrm{g}}^2}{1+\epsilon}\biggl\{\left(\frac{1+\epsilon}{\tstop}\right)^2Dk^2+\frac{1+\epsilon}{\tstop}\cd^2k^2\notag\\
&+\epsilon\nu k^2\left(2\pi G\Sigma_0k+\frac{Dk^2}{\tstop}\right)\biggr\}+\frac{\epsilon\omega_{\dst}^2}{1+\epsilon}\notag\\
&\times\left\{\frac{1+\epsilon}{\tstop}\cs^2k^2+\nu k^2\left(3\Omega^2+\cs^2k^2-2\pi G\Sigma_0k\right)\right\}\notag\\
&+\frac{\epsilon\nu k^2}{\left(1+\epsilon\right)\tstop}\left\{\frac{1+\epsilon}{\tstop}\left(\cd^2-\cs^2\right)k^2+Dk^2\left(\cs^2k^2-\Omega^2\right)\right\}\notag\\
&+\frac{\nu k^2}{1+\epsilon}\left\{\Omega^2+\left(\frac{1+\epsilon}{\tstop}\right)^2\right\}\notag\\
&\times\left\{3\Omega^2+\cs^2k^2-2\pi G\left(1+\epsilon\right)\Sigma_0k\right\},\\
B_0\equiv&\frac{\nu k^2}{1+\epsilon}\left\{\left(\frac{1+\epsilon}{\tstop}\right)^2Dk^2+\frac{1+\epsilon}{\tstop}\cd^2k^2\right\}\notag\\
&\times\left(3\Omega^2+\cs^2k^2-2\pi G\Sigma_0k\right)\notag\\
&-\nu k^2\frac{\epsilon\cs^2k^2}{\tstop}\left(2\pi G\Sigma_0k+\frac{Dk^2}{\tstop}\right)\label{eq:B0}.
\end{align}
We here neglect the second and higher order terms of $\nu k^2$ by assuming that the turbulence is weak ($\alpha\ll 1$). Two modes obtained from Equation (\ref{eq:TVGIapdis}) are the secular GI and TVGI. In fact, if we assume $\nu=0$, the solutions of Equation (\ref{eq:TVGIapdis}) gives the static mode ($n=0$) and the secular GI mode obtained from Equation (\ref{eq:NEWapdis}). Figure \ref{fig:Q5ep01taus03alp1e-3_dispersion} shows the dispersion relations of TVGI and the secular GI for $\alpha=10^{-3}, \epsilon=0.1,\taus=0.3$ and $Q=5$. We calculate the diffusion coefficient and the velocity dispersion from Equations (\ref{YLdiff}) and (\ref{YLcd}). In this case, the secular GI does not grow. Since $n$ is real for both TVGI and the secular GI, they are not oscillating modes. We find that TVGI can grow even in the case where the secular GI is stable. The growth rate of TVGI is very small at long wavelengths because the angular momentum transport by the viscosity becomes ineffective as $k$ decreases. We derive the condition for the growth of TVGI from the approximate dispersion relation (Equation (\ref{eq:TVGIapdis})) in the case where the secular GI is stable. We assume that the disk is self-gravitationally stable and $\omega_{\mathrm{gd}}^2>0$, that is,
\begin{equation}
\frac{\left(1+\epsilon\right)^{3/2}}{\sqrt{1+\epsilon\left(\cd/\cs\right)^2}}<Q.
\end{equation}
We consider cases with $B_2>0$ below, which is satisfied when $\omega_{\mathrm{gd}}^2>0$ and $\epsilon\nu k^2/\tstop\ll\Omega^2$. In this case, the condition is that for a certain wavenumber $k$ Equation (\ref{eq:TVGIapdis}) has one negative solution and one positive solution, which is equivalent to the condition for the existence of the wavenumber where $B_0<0$ is satisfied. From Equation (\ref{eq:B0}), we obtain the following quadratic equation for $k>0$:
\begin{align}
\frac{B_0\tstop}{\nu k^4}=&3\left(\frac{1+\epsilon}{\tstop}D+\cd^2\right)\Omega^2\notag\\
&-2\pi G\Sigma_0\left(\frac{1+\epsilon}{\tstop}D+\cd^2+\epsilon\cs^2\right)k\notag\\
&+\left(\frac{D}{\tstop}+\cd^2\right)\cs^2k^2.\label{eq:appB0TVGI}
\end{align}
For the existence of wavenumbers that satisfy $B_0<0$, the discriminant of the right hand side of Equation (\ref{eq:appB0TVGI}) must be positive. We then find the following condition for the instability: 
\begin{figure}
	\begin{center}
		\includegraphics[width=\columnwidth]{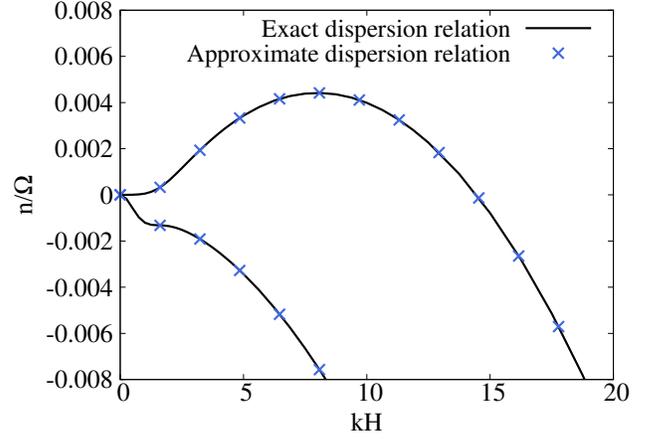}
	\end{center}
    		\caption{Dispersion relations of TVGI and the secular GI for $\alpha=10^{-4}, \epsilon=0.1,\taus=0.03$ and $Q=4$. The horizontal axis is the normalized wavenumber, and the vertical axis is the normalized growth rate $n$. The solid lines represent the exact dispersion relation. The blue cross mark is the approximated dispersion relation (Equation (\ref{eq:TVGIapdis})). In this case, both TVGI and the secular GI grow.}
   		 \label{fig:Q4ep01taus003alp1e-4_dispersion}
\end{figure}
\begin{figure*}
	\begin{center}
		\includegraphics[width=1.5\columnwidth]{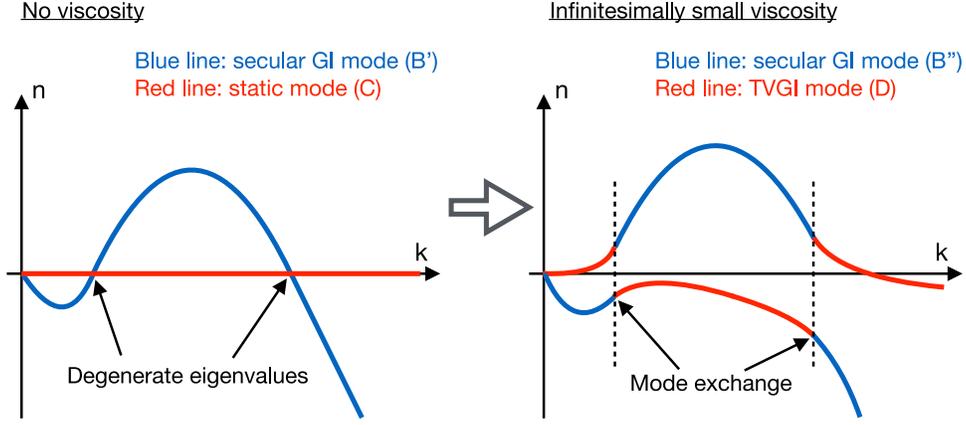}
	\end{center}
    		\caption{Schematic picture that shows how the mode exchange between TVGI and the secular GI occurs by the turbulent viscosity. We show the dispersion relations of the secular GI and the statice mode obtained without the viscosity on the left figure, and those obtained with the viscosity on the right figure. The blue line shows the secular GI mode, and the red line shows the static mode or TVGI. The labels (B'), (B''), (C) and (D) shown in the legends correspond to the labels shown in Figure \ref{fig:modeclass}. The mode exchange occurs at the wavelengths where the eigenvalue and eigenfunction degenerate for the case without the turbulent viscosity.}
   		 \label{fig:mode_exchange}
\end{figure*}
\begin{equation}
\frac{3Q^2\left(\tstop\cd^2+D\right)\left[\tstop\cd^2+\left(1+\epsilon\right)D\right]}{\left[\tstop\left(\cd^2+\epsilon\cs^2\right)+\left(1+\epsilon\right)D\right]^2}<1.\label{eq:app_unstcond_TVGI}
\end{equation}
The left hand side of Euqation (\ref{eq:app_unstcond_TVGI}) does not depend on $\nu$, which is partly because we assume the turbulence is weak. Another reason is that the infinitesimally small viscosity is enough for TVGI to grow. To clarify the physical meaning of the above condition, we only consider the leading term in Equation (\ref{eq:app_unstcond_TVGI}) by assuming $\alpha\ll\taus\ll 1$ and $\cd^2/\cs^2\sim D\Omega/\cs^2\sim\alpha$, and obtain
\begin{equation}
3\left(1+\epsilon\right)\left(\frac{QD}{\epsilon\tstop\cs^2}\right)^2\lesssim 1,\label{eq:easy_unstcond_TVGI}
\end{equation}
or
\begin{equation}
\left(\tstop\pi G\Sigma_0 H^{-1}\right)^{-1}\lesssim\frac{\left(DH^{-2}\right)^{-1}}{\sqrt{3\left(1+\epsilon\right)}}.\label{eq:unstcond_comparetime}
\end{equation}
The left hand side of Equation (\ref{eq:unstcond_comparetime}) represents a timescale in which the dust transverses the length $H=\cs\Omega^{-1}$ with the terminal velocity. The right hand side represents a diffusive time scale of the surface density perturbation of the dust with the length scale $\sim H$. Thus, Equation (\ref{eq:app_unstcond_TVGI}) represents the condition that the dust grains accumulate with the terminal velocity by overcoming the turbulent diffusion. This physical picture is analogous to that of the one-component secular GI discussed in \citet{Youdin2011}. We can also estimate the most unstable wavelength when the higher order terms of $\nu k^2$ are negligible. The growth rate of TVGI is also small in this limit since it is determined by the efficiency of the angular momentum transport by the turbulent viscosity. Hence, in this case, the growth rate of TVGI is approximately given by $-B_0/B_1$ (see Equation (\ref{eq:TVGIapdis})). By neglecting the higher order terms of $\nu k^2$, the growth rate is reduced as follows:
\begin{align}
n\simeq &-\nu k^2\bigl[3\Omega^2D\left(1+\epsilon\right)\notag\\
&-2\pi G\Sigma_0\left\{\tstop\epsilon\cs^2+D\left(1+\epsilon\right)\right\}k+D\cs^2k^2\bigr]\notag\\
&\times\bigl[\left\{\tstop\epsilon\cs^2+D\left(1+\epsilon\right)\right\}\Omega^2\notag\\
&-2\pi G\left(1+\epsilon\right)\Sigma_0\left\{\tstop\epsilon\cs^2+D\left(1+\epsilon\right)\right\}k\notag\\
&+D\left(1+\epsilon\right)\cs^2k^2\bigr]^{-1}\label{eq:app_n_TVGI_mst_unst}
\end{align}
We here neglect $\cd^2k^2$ since this term has a smaller effect to stabilize TVGI than the diffusion. The most unstable wavenumber $k_{\mathrm{max}}$ is of the order of a wavenumber where $n/\nu k^2$ has the local maximum, which is  
\begin{align}
k_{\max}&\sim\frac{\pi G\Sigma_0\left[\tstop\epsilon\cs^2+D\left(1+\epsilon\right)\right]}{D\cs^2}\notag\\
&=\frac{1+\epsilon}{Q}H^{-1}+\frac{\epsilon\taus}{\left(D\Omega\cs^{-2}\right)Q}H^{-1}.\label{eq:app_mst_unst_k}
\end{align}
The right hand side of the Equation (\ref{eq:app_mst_unst_k}) is about $4.5H^{-1}$ for $\alpha=10^{-3}, \epsilon=0.1,\taus=0.3$ and $Q=5$, which is consistent with the most unstable wavenumber seen in Figure \ref{fig:Q5ep01taus03alp1e-3_dispersion}. We note that when the higher order terms of $\nu k^2$ are not negligible, the condition for the instability depends on $\nu$ although Equations (\ref{eq:app_unstcond_TVGI}) and (\ref{eq:unstcond_comparetime}) do not.
\begin{figure}
	\begin{center}
		\includegraphics[width=\columnwidth]{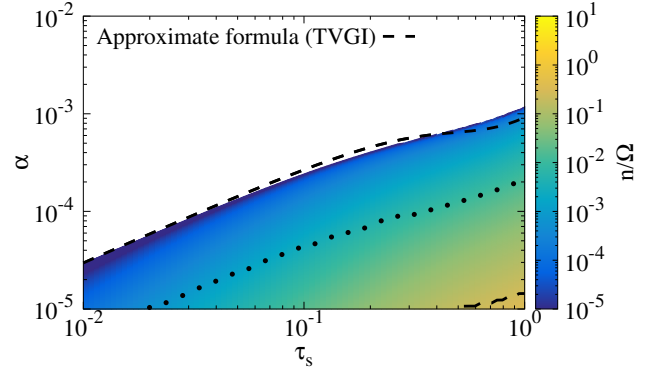}
	\end{center}
    		\caption{Maximum growth rate of the instabilities for $\epsilon=0.05$ and $Q=10$. The horizontal axis is the normalized stopping time $\taus$, and the vertical axis is the strength of turbulence $\alpha$. The color represents the maximum growth rate normalized by $\Omega$. The short-dashed line represents the condition for the growth of TVGI (Equation (\ref{eq:app_unstcond_TVGI})). The most unstable mode is TVGI in the colored region above the dotted line. In the region between the dotted line and the long-dashed line, the secular GI is the fastest growing mode. The dust GI mode is the most unstable mode below the long-dashed line as in Figure \ref{fig:Q3ep01-growthrate_wo_thickness_wo_viscosity}. All of those instabilities are stable in the white region.}
   		 \label{fig:Q10ep005-growthrate_w_viscosity_wo_thickness}
\end{figure}

\begin{figure*}
	\begin{tabular}{c}
		\begin{minipage}{0.5\hsize}
			\begin{center}
				\includegraphics[width=0.9\columnwidth]{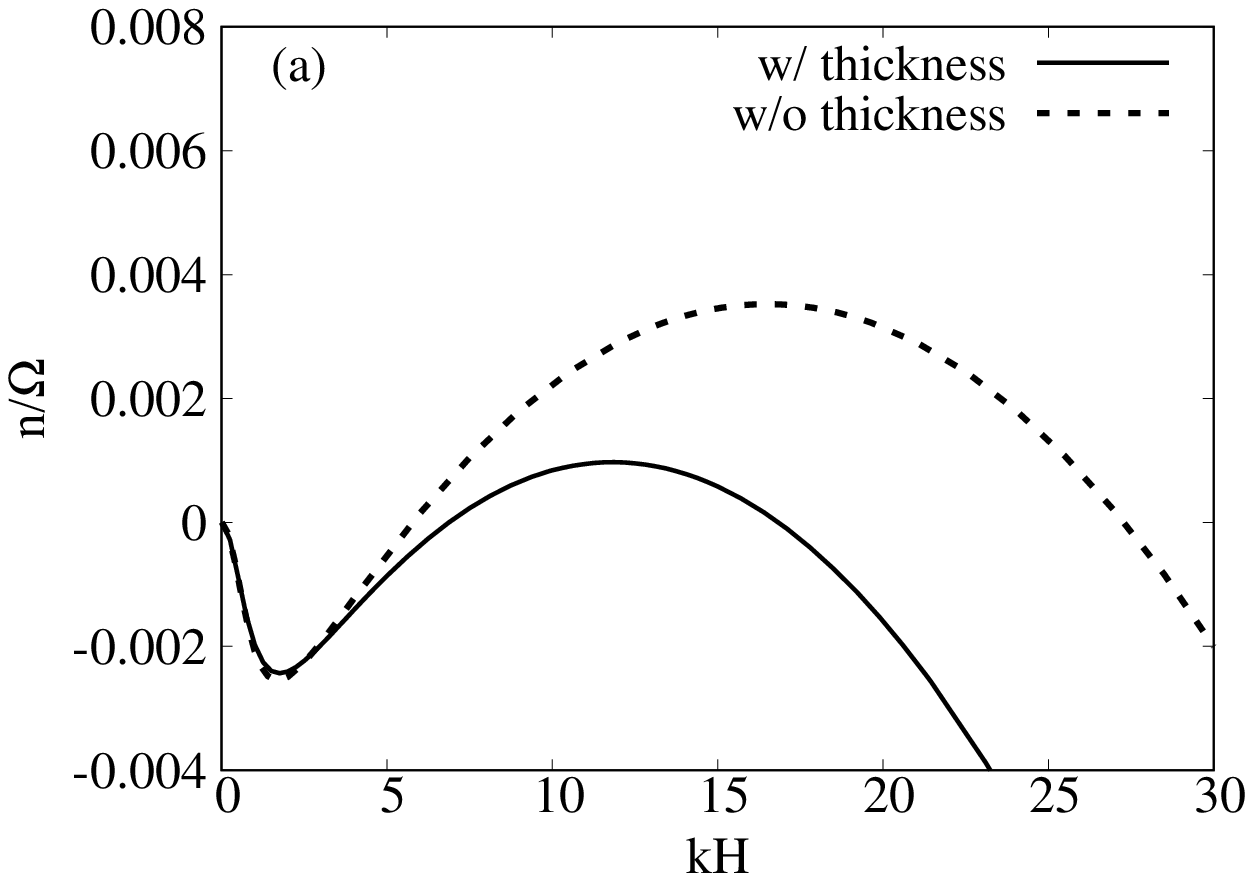}
			\end{center}
		\end{minipage}
		\begin{minipage}{0.5\hsize}
			\begin{center}
				\includegraphics[width=0.9\columnwidth]{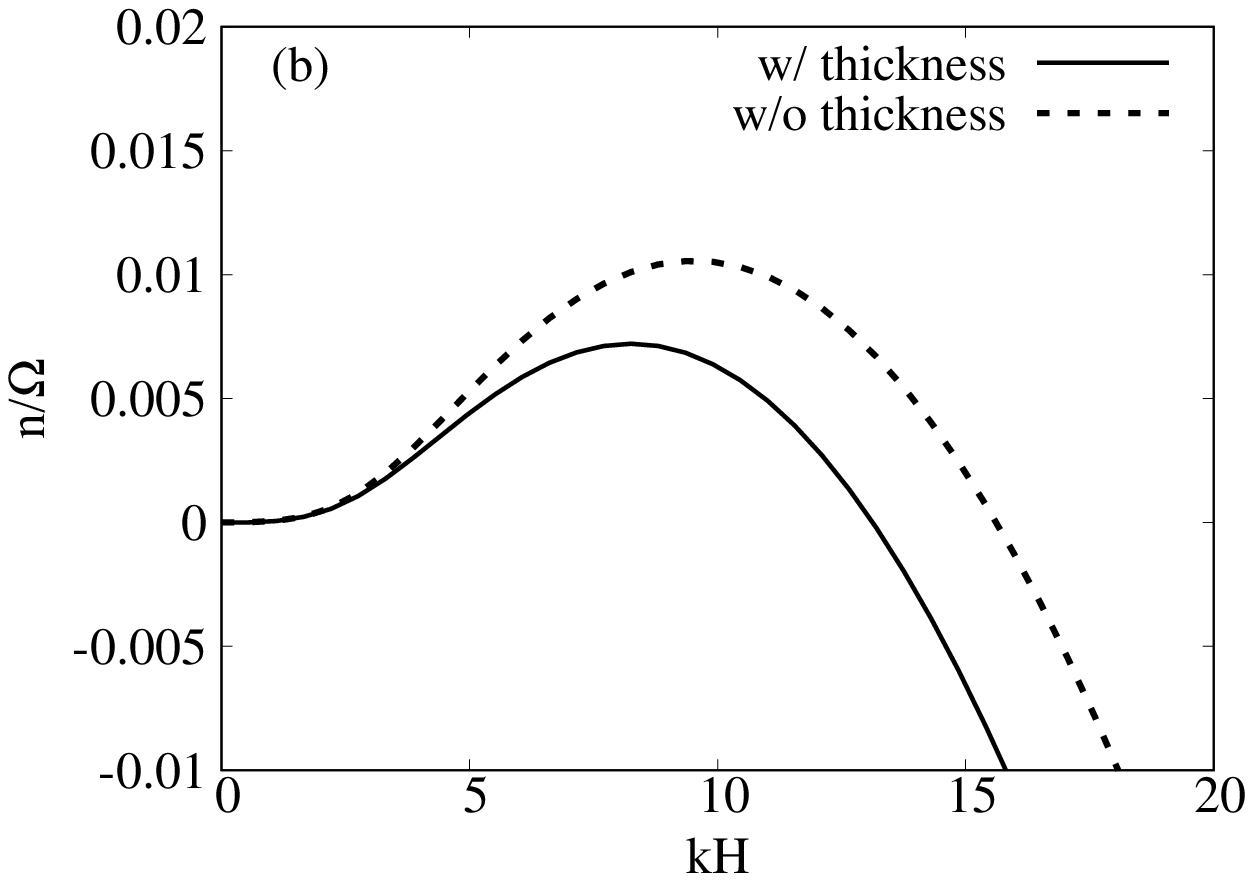}
			\end{center}
		\end{minipage}
	\end{tabular}
\caption{Dispersion relations of the secular GI (left panel) and TVGI (right panel) for $\epsilon=0.05$ and $Q=10$. The normalized stopping time and the strength of turbulence ($\taus,\alpha$) are set to be ($0.1,2.5\times10^{-5}$) for the left panel and ($1,2.5\times10^{-4}$) for the right panel (see also, Figure \ref{fig:Q10ep005-growthrate_w_viscosity_w_thickness}). We neglect the turbulent viscosity on gas for the left panel so that only the secular GI grows. The horizontal axis is the normalized wavenumber, and the vertical axis is the normalized growth rate. The solid lines represent the dispersion relation for a disk with finite thickness. The dashed lines are the dispersion relation for a razor thin disk. The growth rate and the most unstable wavenumber become smaller because of the decrease of the self-gravity due to the effect of the disk thickness.}
 \label{fig:eff_of_thickness}
\end{figure*}

Next, we consider the other case where the secular GI is also unstable. Figure \ref{fig:Q4ep01taus003alp1e-4_dispersion} shows the dispersion relations for $\alpha=10^{-4}, \epsilon=0.1,\taus=0.03$ and $Q=3$. Although TVGI and the secular GI are unstable in this case (Equations (\ref{eq:app_unstcond_NEWSGI}) and (\ref{eq:app_unstcond_TVGI})), only one growing mode appears. This is because the mode exchange occurs between TVGI and the secular GI (Figure \ref{fig:mode_exchange}). If we neglect the viscosity, eigenvalues and eigenfunctions of the static mode and the secular GI mode degenerate at wavenumbers where the growth rate of the secular GI is zero. If we include the finite viscosity, the dispersion relations of the destabilized static mode (TVGI) and the secular GI reconnect at the wavenumbers, which results in one growing mode. As shown in Figure \ref{fig:mode_exchange}, at wavenumbers where the secular GI is unstable for $\nu=0$, we designate the growing mode as the secular GI even when $\nu$ is not zero. On the other hand, in a band of wavenumber $k$ where the secular GI is stable for $\nu=0$, we call the mode TVGI. Figure \ref{fig:Q10ep005-growthrate_w_viscosity_wo_thickness} shows the maximum growth rate of the instabilities for $\epsilon=0.05$ and $Q=10$. In the colored region above the dotted line, TVGI is the fastest growing mode. The dotted line almost coincides with the maximum $\alpha$ determined by Equation (\ref{eq:app_unstcond_NEWSGI}). The secular GI is the fastest growing mode in the region enclosed by the dotted line and the long-dashed line. Figure \ref{fig:Q10ep005-growthrate_w_viscosity_wo_thickness} also shows that, compared to the secular GI, TVGI can grow in a region where the turbulence and the friction are strong. We expect that TVGI grows earlier than the secular GI since the stopping time becomes larger as the dust grows in protoplanetary disks (see Figure \ref{fig:Q10ep005-growthrate_w_viscosity_wo_thickness}). As described above, the dust grains accumulate through the growth of TVGI. Therefore, TVGI should be a promising mechanism to form planetesimals.

The turbulent viscosity makes two of the modified density waves overstable \citep[cf.,][]{Schmit1995}. We do not discuss the properties of this viscous overstability in this work since the growth time is found to be longer than the typical disk lifetime (see also Section \ref{sec:ring_disk}).

\section{Discussion}\label{sec:discussion}
\subsection{Effect of Disk Thickness}\label{sec:analysis_dustsubdisk}
\begin{figure}
	\begin{center}
		\includegraphics[width=\columnwidth]{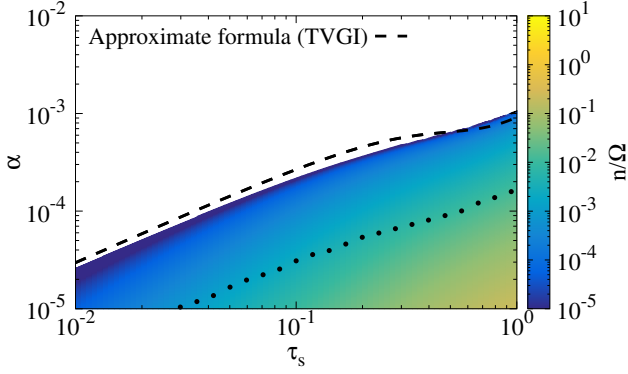}
	\end{center}
    		\caption{Maximum growth rate obtained with the effect of the disk thickness. The $Q$ and $\epsilon$ are set to be the same value as those of Figure \ref{fig:Q10ep005-growthrate_w_viscosity_wo_thickness}. The horizontal axis is the normalized stopping time $\taus$, and the vertical axis is the strength of turbulence $\alpha$. The color represents the maximum growth rate normalized by $\Omega$. The short-dashed line is the same as that shown in Figure \ref{fig:Q10ep005-growthrate_w_viscosity_wo_thickness}. The most unstable mode is TVGI in the colored region above the dotted line. In the region below the dotted line, the secular GI is the fastest growing mode.  Both instabilities are stable in the white region.}
   		 \label{fig:Q10ep005-growthrate_w_viscosity_w_thickness}
\end{figure}

We assume that the disk is razor thin in the above analysis, while a real disk has a finite thickness. The thickness of the disk reduces the self-gravity estimated for the infinitesimally thin disk case. We here show the effect on the dispersion relation and the maximum growth rate. The self-gravitational potential $\delta\Phi$ reduced by the disk thickness is approximately given as follows \citep{Vandervoort1970,Shu1984}:
\begin{equation}
\delta\Phi=-\frac{2\pi G}{k}\left(\frac{\delta\Sigma}{1+kH}+\frac{\delta\sigmad}{1+k\hd}\right),
\end{equation}
where $\hd$ is the thickness of the dust disk that is represented by the following \citep{YL2007}:
\begin{equation}
\hd=H\left(1+\frac{\taus}{\alpha}\frac{1+2\taus}{1+\taus}\right)^{-1/2}.\label{eq:hd_h}
\end{equation}
Figure \ref{fig:eff_of_thickness} shows how the dispersion relations of the secular GI and TVGI change due to the thickness for $\epsilon=0.05$ and $Q=10$. The normalized stopping time and the strength of turbulence ($\taus,\alpha$) are set to be ($0.1,2.5\times10^{-5}$) for the left panel of Figure \ref{fig:eff_of_thickness}, while the right panel is for  ($\taus,\alpha$)$=$($1,2.5\times10^{-4}$). The most unstable mode is the secular GI and TVGI for the former and latter cases, respectively. The growth rate decreases by the effect of the disk thickness. Figure \ref{fig:Q10ep005-growthrate_w_viscosity_w_thickness} shows the maximum growth rate as a function of $\taus$ and $\alpha$ for the same parameters as in Figure \ref{fig:Q10ep005-growthrate_w_viscosity_wo_thickness}. Although the maximum growth rate is smaller than that in Figure \ref{fig:Q10ep005-growthrate_w_viscosity_wo_thickness}, the extent of the unstable region is almost the same, and the maximum $\alpha$ for a certain stopping time $\taus$ does not change more than a factor of two. 

The dust disk is generally thinner than the gaseous disk (see Equation (\ref{eq:hd_h})). In reality, the gas above the dust disk does not interact with the dust grains through the friction, while in the above analysis on the infinitesimally thin disk the back reaction is assumed to be exerted on all of the gas. We need to exclude the gas located above the dust disk from our analysis \citep[see also,][]{Latter2017}. In this subsection, we simply assume that the vertical density profile is given by the following Gaussian function and investigate stability in the dust disk:
\begin{equation}
\rho_0\equiv\frac{\Sigma_0}{\sqrt{2\pi}H}\exp\left(-\frac{z^2}{2H^2}\right),
\end{equation}
\begin{equation}
\rho_{\dst,0}\equiv\frac{\sigmadup}{\sqrt{2\pi}\hd}\exp\left(-\frac{z^2}{2\hd^2}\right),
\end{equation}
where $\rho_0$ and $\rho_{\dst,0}$ are the mass density of the gas and the dust, respectively. The vertical distance from the disk midplane is denoted by $z$. In the following analysis, we use the surface density obtained from the vertical integration of those gas and dust densities. There is a large uncertainty in the appropriate range of the vertical integration. Here, we integrate these densities in $-3\hd\leq z \leq 3\hd$. Although the dust density at $z=3\hd$ is about 0.01 times smaller than that at the midplane, the dust-to-gas mass ratio is still large because the dust scale height is much smaller than the gas scale height and the dust concentrates in the midplane. In addition, the strength of the friction on the dust is independent from the dust density, which means that the dust even at a low density region can be unstable as a result of the friction. The vertical integration of the mass densities gives
\begin{equation}
\Sigma_0'\equiv\int_{-3\hd}^{3\hd}\rho_0 dz=\Sigma_0\mathrm{erf}\left(\frac{3\hd}{\sqrt{2}H}\right),
\end{equation}
\begin{equation}
\sigmadup'\equiv\int_{-3\hd}^{3\hd}\rho_{\dst,0} dz=\sigmadup\mathrm{erf}\left(\frac{3}{\sqrt{2}}\right)\simeq0.997\sigmadup,
\end{equation}
where $\mathrm{erf}(x)$ is the error function. The relation between the mid-plane dust-to-gas mass ratio and $\sigmadup'/\Sigma_0'$ is
\begin{equation}
\frac{\rho_{\dst,0}(z=0)}{\rho_0(z=0)}=\frac{\sigmadup' H}{\Sigma_0'\hd}\mathrm{erf}\left(\frac{3\hd}{\sqrt{2}H}\right)\left[\mathrm{erf}\left(\frac{3}{\sqrt{2}}\right)\right]^{-1}.
\end{equation}
We perform the linear analysis by using these surface densities and the dust-to-gas mass ratio $\sigmadup'/\Sigma_0'$. In this analysis, we introduce the modified Toomre parameter for the gas disk $\tilde{Q}\equiv\cs\Omega/\pi G\Sigma_0'$. We modify the self-gravitational potential by using $\hd$ for both gas and dust:
\begin{equation}
\delta\Phi=-\frac{2\pi G}{k}\frac{\delta\Sigma'+\delta\sigmad'}{1+k\hd}.
\end{equation}
\begin{figure}
	\begin{center}
		\includegraphics[width=\columnwidth]{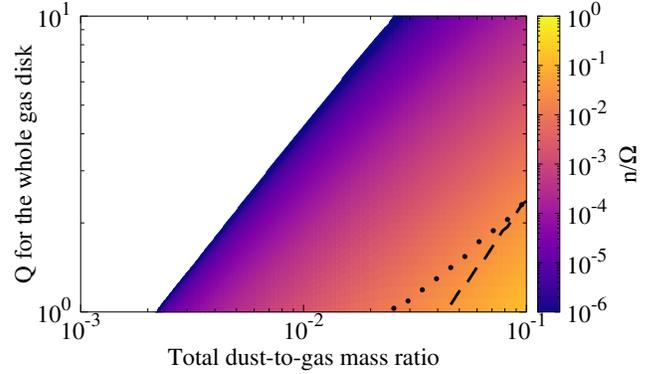}
	\end{center}
    		\caption{Maximum growth rate of the instabilities for $\taus=0.1$ and $\alpha=10^{-4}$. The horizontal axis is the total dust-to-gas mass ratio $\sigmadup/\Sigma_0$. The vertical axis is Toomre's $Q$ value for the whole gas disk, that is, $Q=\cs\Omega/\pi G\Sigma_0$. The color represents the maximum growth rate normalized by the angular velocity $\Omega$. In the colored region above the dotted line, TVGI is the most unstable mode. The secular GI is the fastest growing mode in the region enclosed by the dotted and dashed lines. The dust GI mode becomes the most unstable mode below the dashed line.}
   		 \label{fig:taus01alph1e-4-growthrate_w_viscosity_w_thickness}
\end{figure}
Figure \ref{fig:taus01alph1e-4-growthrate_w_viscosity_w_thickness} shows the maximum growth rate as a function of total dust-to-gas mass ratio $\sigmadup/\Sigma_0$ and $Q=\cs\Omega/\pi G\Sigma_0$. The normalized stopping time $\taus$ and the strength of the turbulence $\alpha$ are set to be $0.1$ and $10^{-4}$, respectively. In this case, $\hd/H$ is about 0.03, and, thus, $\tilde{Q}/Q$ is about 14, meaning that $\tilde{Q}$ is larger than 10 in the whole parameter space shown in Figure \ref{fig:taus01alph1e-4-growthrate_w_viscosity_w_thickness}. The dust-to-gas mass ratio in the dust disk $\sigmadup'/\Sigma_0'$ is also about ten times larger than $\sigmadup/\Sigma_0$. The secular GI becomes the most unstable mode only in very small parameter space, which is because $\tilde{Q}$ is too large and the dust GI mode is unstable enough to grow faster than the secular GI. On the other hand, TVGI can grow in much larger parameter space. Since the self-gravity of the dust is important for the growth, TVGI grows even for large $\tilde{Q}$. Therefore, we expect that TVGI operates in the dust disk even if we consider the vertical structure. The quantitative results do not change even if we vary the range of the vertical integration. In Appendix \ref{ap:dustsubdisk}, we show results for cases where we integrate the densities in $-\hd\leq z \leq \hd$ and $-2\hd\leq z \leq 2\hd$. We note that gas above the dust disk, which is free from the frictional force, would affect the motion in the dust disk through the gravitational interaction. Thus, we might underestimate the self-gravity in this analysis. The parameter space unstable to TVGI and the secular GI might become larger if we evaluate the self-gravity more precisely. In order to examine the effect of the upper gas on the instabilities in the dust disk, we need to perform the multidimensional analysis, which is beyond the scope of this paper. Moreover, the dust is expected to be less diffusive in high dust-to-gas ratio regions. This was also reported by \citet[][]{Schreiber2018}, in which they performed numerical simulations on the streaming instability. The less radial diffusivity makes the disk more unstable so that the instabilities grow in the larger parameter space than that shown in Figure \ref{fig:taus01alph1e-4-growthrate_w_viscosity_w_thickness}.

\subsection{Ring Formation through TVGI}\label{sec:ring_disk}
The secular GI has been proposed as one of the possible mechanisms to create multiple ring-like structures recently observed in some protoplanetary disks. \citet{Takahashi2016} performed the linear analysis by adopting physical values obtained from the observation of HL Tau \citep{ALMA-Partnership2015} as the unperturbed state and showed that the multiple rings observed in the HL Tau disk can form via the secular GI. TVGI also has potential to be the mechanism to form multiple rings because dust grains accumulate in the radial direction through the growth of TVGI. Here, we show results of the linear analysis based on one of the disk model used in \citet{Takahashi2016} and discuss where TVGI operates and creates rings in the HL Tau disk.

\begin{figure*}
	\begin{tabular}{c}
		\begin{minipage}{0.5\hsize}
			\begin{center}
				\includegraphics[width=0.9\columnwidth]{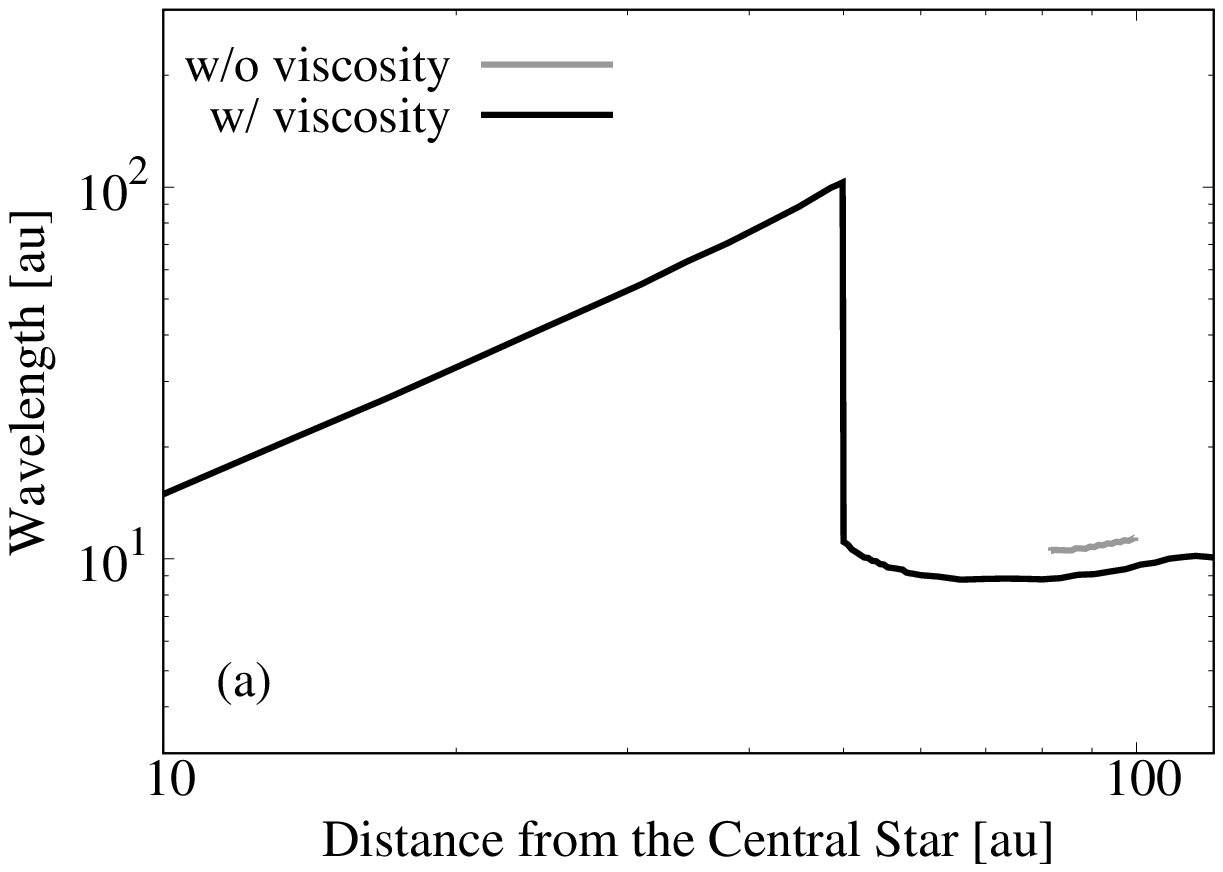}
			\end{center}
		\end{minipage}
		\begin{minipage}{0.5\hsize}
			\begin{center}
				\includegraphics[width=0.9\columnwidth]{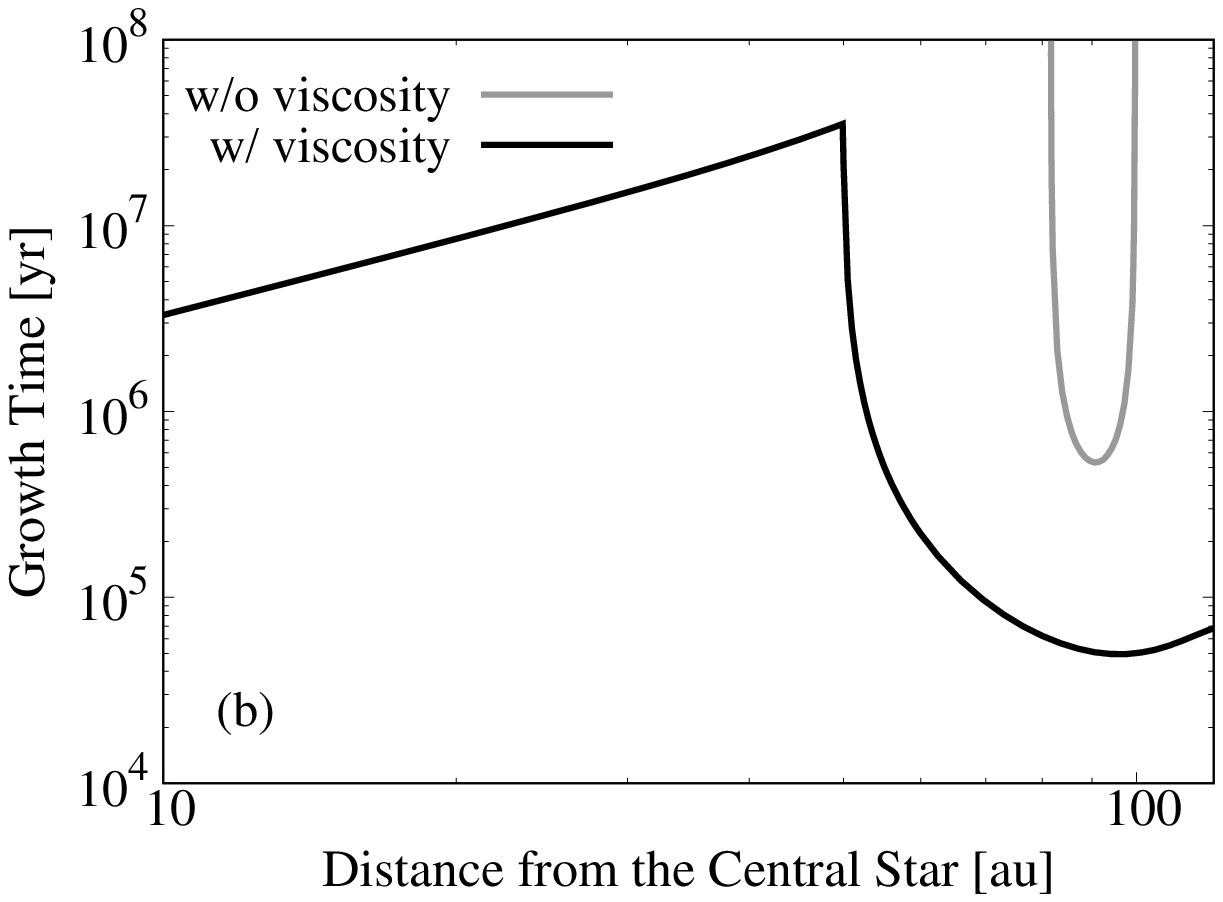}
			\end{center}
		\end{minipage}
	\end{tabular}
\caption{The most unstable wavelength of the instabilities (left panel) and its growth time (right panel) obtained by adopting the exponentially cutoff disk model. The horizontal axis of both panel is the radial distance from the central star. The black and gray lines show the results obtained with and without the turbulent viscosity, respectively. Only the secular GI grows when we neglect the turbulent viscosity. The spatial extent of the region unstable to the secular GI is $80\;\mathrm{au}\lesssim r \lesssim 100\;\mathrm{au}$. When we consider the effect of the viscosity,  there are three growing mode: the secular GI, TVGI, and the viscous overstablity. TVGI grows at $50\;\mathrm{au}\lesssim r\lesssim 80\;\mathrm{au}$ and $r\gtrsim100\;\mathrm{au}$. Although the viscous overstablity grows in the inner region ($r\lesssim 50\;\mathrm{au}$), the growth time is larger than the typical disk life time. In addition, the most unstable wavelength of the viscous overstability is larger than the radius. We thus do not discuss the viscous overstability in this paper.}
 \label{fig:HLTau_exponentially_cut_off}
\end{figure*}

We use the following dust surface density profile and gas temperature profile that reflect the observation of HL Tau \citep{Pinte2016,Kwon2015}, referred to as the ``exponentially cutoff disk model" in \citet{Takahashi2016}:
\begin{equation}
\sigmad(r)=0.51\left(\frac{r}{R_{\mathrm{c}}}\right)^{-\gamma}\exp\left[-\left(\frac{r}{R_{\mathrm{c}}}\right)^{2-\gamma}\right]\;\;\;\;\mathrm{[g \;cm}^{-2}],
\end{equation}
\begin{equation}
T(r)=30\left(\frac{r}{20\;\mathrm{[au]}}\right)^{-0.65}\;\;\;\;\mathrm{[K]},
\end{equation}
where $R_{\mathrm{c}}=80.2$ au, $\gamma=-0.2$. We assume that the mass of the central star is $1\msun$, and the dust size and the internal density are 3 mm and 3 $\mathrm{g \;cm}^{-3}$, respectively. The dust-to-gas mass ratio is set to be 0.02 as in \citet{Takahashi2016} \citep[see also,][]{Kwon2015}. The strength of the turbulence $\alpha$ is assumed to be $3\times 10^{-4}$. Such weak turbulence reflects the observed very thin dust disk \citep{Pinte2016}. Figure \ref{fig:HLTau_exponentially_cut_off} shows the most unstable wavelength and its growth time $n^{-1}$. In order to compare the extent of the regions unstable to the secular GI and/or TVGI, we show the results of the linear analysis with and without the turbulent viscosity. Only the secular GI grows when we do not consider the viscosity, and it forms multiple rings in the region $80\;\mathrm{au}\lesssim r \lesssim 100\;\mathrm{au}$. When we include the viscosity in the equations, there exist three instabilities: the secular GI, TVGI and the viscous overstability \citep[cf.,][]{Schmit1995}. The region where the viscous overstability is the most unstable mode is $r\lesssim 50\;\mathrm{au}$. We do not discuss the viscous overstability since the most unstable wavelength of the viscous overstability is larger than the radius and its growth time is longer than the typical disk lifetime. The regions where TVGI is the most unstable mode are $50\;\mathrm{au}\lesssim r\lesssim 80\;\mathrm{au}$ and $r\gtrsim100\;\mathrm{au}$. As shown in Figure \ref{fig:HLTau_exponentially_cut_off}, the most unstable wavelength is about 10 au, which is almost independent of the radius. This wavelength is comparable to the width of the rings observed in the HL Tau disk. As discussed in \citet{Takahashi2016}, the growth timescale is required to be less than $10^6$ yr since HL Tau is thought to be young. We find that the requirement on the timescale is satisfied in $r\gtrsim 53\;\mathrm{au}$. Therefore, we conclude that TVGI can form some of the multiple rings observed in HL Tau. We also find that TVGI grows in the larger extent of the disk than the secular GI.

Recent observations with ALMA have found ring structures in other disks \citep[e.g.,][]{Andrews2016,Isella2016,Sheehan2017,Fedele2018, Andrews2018}. The Disk Substructures at High Angular Resolution Project \citep[DSHARP,][]{Andrews2018} observed 20 disks with high spatial resolution ($\sim 5$ au) and found that most of them have rings and gaps although non-axisymmetric structures were also observed \citep[see also, ][]{Huang2018}. According to \citet{Dullemond2018}, some of the observed rings have width comparable to the gas scale height or smaller. Such relatively narrow rings can form via TVGI. Equation (\ref{eq:app_mst_unst_k}) gives the most unstable wavelength $\lambda_{\mathrm{max}}$ normalized by the gas scale height:
\begin{equation}
\frac{\lambda_{\mathrm{max}}}{H}\equiv\frac{2\pi}{k_{\mathrm{max}}H}\sim\frac{2\pi Q\left(D\Omega\cs^{-2}\right)}{\epsilon\taus+\left(1+\epsilon\right)\left(D\Omega\cs^{-2}\right).}
\end{equation}
For $D\Omega\cs^{-2}\sim\alpha\ll\taus$, $\lambda_{\mathrm{max}}/H$ is roughly given by
\begin{equation}
\frac{\lambda_{\mathrm{max}}}{H}\sim\frac{2\pi Q\left(D\Omega\cs^{-2}\right)}{\epsilon\taus}\sim\frac{2\pi Q\alpha}{\epsilon\taus}. \label{eq:lambda_max}
\end{equation}
Equation (\ref{eq:easy_unstcond_TVGI}) determines the upper limit on the turbulent strength $\alpha$ for the growth of TVGI as follows:
\begin{equation}
\alpha\lesssim\frac{1}{\sqrt{3\left(1+\epsilon\right)}}\frac{\epsilon\taus}{Q}\sim\frac{1}{\sqrt{3}}\frac{\epsilon\taus}{Q},\label{eq:ulimit_alpha}
\end{equation}
where $\epsilon\ll 1$ is assumed. From Equations (\ref{eq:lambda_max}) and (\ref{eq:ulimit_alpha}), we obtain the upper limit on the maximum wavelength independent from any parameters:
\begin{equation}
\frac{\lambda_{\mathrm{max}}}{H} \lesssim \frac{2\pi}{\sqrt{3}}\simeq 3.6.\label{eq:ulimit_lambda}
\end{equation}
The most unstable wavelength is observable as the distance between peaks of adjacent rings. We note that \citet{Dullemond2018} fitted ring structures by using a Gaussian intensity profile and defined ring widths as standard deviation, which is different from the length that $\lambda_{\mathrm{max}}$ represents. The standard deviation should be smaller than $\lambda_{\mathrm{max}}$ if the observed rings formed via TVGI, and this is consistent with the narrowness of the observed rings. We therefore conclude that TVGI can be the origin of the observed narrow ring structures.

\section{Conclusions}\label{sec:conclusion}
In this work, we re-formulate the equations of gas and dust in a protoplanetary disk in which the turbulence of the gas diffuses dust grains, based on the Reynolds averaging. The equations guarantee the conservation of the total angular momentum of the disk, which is not conserved in the equations used in previous studies. By using the newly formulated equations, we investigate the linear growth of the secular GI. We find that the secular GI is a monotonically growing mode while the secular GI found in the previous work  is overstable depending on parameters. The appearance of the overstability is ascribed to the non-conservation of the total angular momentum. We also find the new instability referred to as TVGI if the turbulent viscosity is considered. TVGI grows in regions where $\alpha$ is larger and $\taus$ is smaller compared with the secular GI. Since the normalized stopping time $\taus$ increases as a dust grain grows, we can expect that TVGI operates in protoplanetary disks earlier than the secular GI. 

We investigate the ring formation in the HL Tau disk via TVGI using the result of the linear analysis. Assuming that the dust size and the strength of turbulence $\alpha$ is constant in the radial direction, we show TVGI can form the observed multiple rings at $50\;\mathrm{au}\lesssim r\lesssim 80\;\mathrm{au}$ and $r\gtrsim100\;\mathrm{au}$. For $80\;\mathrm{au}\lesssim r\lesssim 100\;\mathrm{au}$, the secular GI grows faster than TVGI. We expect that the relatively narrow rings observed recently form through TVGI (see Equation (\ref{eq:ulimit_lambda})). TVGI is also a promising mechanism to form planetesimals since the dust accumulates during the growth of TVGI. It is important to investigate non-linear growth and to discuss the planetesimal formation via TVGI, which is our future work.

\acknowledgments

We thank Shugo Michikoshi and Hiroshi Kobayashi for fruitful discussion. We also thank the anonymous referee for insightful and constructive comments that helped to improve the manuscript. This work was supported by JSPS KAKENHI Grant Number JP18J20360, 16H02160, 18H05436, 18H05437 and 19K14764, and in part by the National Astronomical Observatory of Japan (NAOJ) Atacama Large Millimeter/submillimeter Array (ALMA) Scientific Research Grant Numbers 2016-02A (SZT) .

%




\appendix

\section{Formulation of basic equations with turbulent diffusion}\label{ap:formulation}

In this appendix, we summarize the derivation of the phenomenological equations for the dust. We use the Reynolds averaging in this work. The Reynolds averaging is one of the techniques to analyze turbulence by focusing on longer-term evolution of systems than the typical time scale of the turbulence. We decompose a physical variable $A$ into the time-averaged term $\left<A\right>$ and the short-term fluctuation due to the turbulence $\Delta A\equiv A-\left<A\right>$, where $\left<\Delta A\right>=0$. We average the following equations for a infinitesimally thin and axisymmetric dust disk:
\begin{equation}
\frac{\partial \sigmad}{\partial t}+\frac{1}{r}\frac{\partial\left(r\sigmad v_r\right)}{\partial r}=0,
\end{equation}
\begin{equation}
\frac{\partial \left(\sigmad v_{r}\right)}{\partial t}+\frac{1}{r}\frac{\partial}{\partial r}\left(r\sigmad v_r^2\right)=\sigmad\frac{v_{\phi}^2}{r}-\sigmad\frac{\partial}{\partial r}\left(\Phi-\frac{GM_{\ast}}{r}\right)-\sigmad\frac{v_{r}-u_{r}}{\tstop},\label{eq:ap:eomrd}
\end{equation}
\begin{equation}
\frac{\partial \left(\sigmad v_{\phi}\right)}{\partial t}+\frac{1}{r}\frac{\partial}{\partial r}\left(r\sigmad v_{\phi}v_r\right)=-\sigmad\frac{v_{\phi}v_r}{r}-\sigmad\frac{v_{\phi}-u_{\phi}}{\tstop}.\label{eq:ap:eomphid}
\end{equation}
First, we average the continuity equation and obtain the following equation that describes the time evolution of the mean surface density:
\begin{equation}
\frac{\partial\left<\sigmad\right>}{\partial t}+\frac{1}{r}\frac{\partial\left(r\left<\sigmad\right>\left<v_r\right>\right)}{\partial r}=-\frac{1}{r}\frac{\partial\left(r\left<\Delta\sigmad\Delta v_r\right>\right)}{\partial r}.
\end{equation}
We model the term $\left<\Delta\sigmad\Delta v_i\right>$, based on the gradient diffusion hypothesis \citep[see also,][]{Cuzzi1993}:
\begin{equation}
\left<\Delta\sigmad\Delta v_r\right>=-D\frac{\partial\left<\sigmad\right>}{\partial r},
\end{equation}
\begin{equation}
\left<\Delta\sigmad\Delta v_{\phi}\right>=-\frac{D}{r}\frac{\partial\left<\sigmad\right>}{\partial \phi}=0.
\end{equation}
The Reynolds-averaged continuity equation is given by
\begin{equation}
\frac{\partial\left<\sigmad\right>}{\partial t}+\frac{1}{r}\frac{\partial\left(r\left<\sigmad\right>\left<v_r\right>\right)}{\partial r}=\frac{1}{r}\frac{\partial}{\partial r}\left(rD\frac{\partial\left<\sigmad\right>}{\partial r}\right).\label{eq:app_diffusion}
\end{equation}
This equation is equivalent to Equation (\ref{eq:eocdust}). Similarly, we average Equations (\ref{eq:ap:eomrd}) and (\ref{eq:ap:eomphid}) and obtain
\begin{align}
\frac{\partial\left(\left<\sigmad\right>\left<v_r\right>\right)}{\partial t}&+\frac{1}{r}\frac{\partial}{\partial r}\left[r\left<\sigmad\right>\left(\left<v_r\right>-\frac{D}{\left<\sigmad\right>}\frac{\partial\left<\sigmad\right>}{\partial r}\right)\left<v_r\right>\right]\notag\\
=&\left<\sigmad\right>\frac{\left<v_{\phi}\right>^2}{r}-\left<\sigmad\right>\frac{\partial}{\partial r}\left(\left<\Phi\right>-\frac{GM_{\ast}}{r}\right)-\left<\Delta\sigmad\frac{\partial\Delta\Phi}{\partial r}\right>-\left<\sigmad\right>\frac{\left<v_r\right>-\left<u_r\right>}{\tstop}+\frac{1}{r}\frac{\partial\left(r\sigma_{rr}\right)}{\partial r}-\frac{\sigma_{\phi\phi}}{r}\notag\\
&+\frac{\partial}{\partial t}\left(D\frac{\partial\left<\sigmad\right>}{\partial r}\right)+\frac{1}{r}\frac{\partial}{\partial r}\left(r\left<v_r\right>D\frac{\partial\left<\sigmad\right>}{\partial r}\right)-\frac{\left<\Delta\sigmad\left(\Delta v_{r}-\Delta u_{r}\right)\right>}{\tstop},\label{eq:ap:momr_av}
\end{align}
\begin{align}
\frac{\partial\left(\left<\sigmad\right>\left<v_{\phi}\right>\right)_t}{\partial t}&+\frac{1}{r}\frac{\partial}{\partial r}\left[r\left<\sigmad\right>\left(\left<v_r\right>-\frac{D}{\left<\sigmad\right>}\frac{\partial\left<\sigmad\right>}{\partial r}\right)\left<v_{\phi}\right>\right]\notag\\
&=-\frac{\left<\sigmad\right>\left<v_{\phi}\right>}{r}\left(\left<v_r\right>-\frac{D}{\left<\sigmad\right>}\frac{\partial\left<\sigmad\right>}{\partial r}\right)-\left<\sigmad\right>\frac{\left<v_{\phi}\right>-\left<u_{\phi}\right>}{\tstop}+\frac{1}{r}\frac{\partial \left(r\sigma_{r\phi}\right)}{\partial r}+\frac{\sigma_{r\phi}}{r}-\frac{\left<\Delta\sigmad\left(\Delta v_{\phi}-\Delta u_{\phi}\right)\right>}{\tstop}.\label{eq:ap:momphi_av}
\end{align}
where
\begin{equation}
\sigma_{rr}\equiv-\left<\sigmad\right>\left<\Delta v_r^2\right>-\left<\Delta\sigmad\Delta v_r^2\right>,
\end{equation}
\begin{equation}
\sigma_{r\phi}\equiv-\left<\sigmad\right>\left<\Delta v_r\Delta v_{\phi}\right>-\left<\Delta\sigmad\Delta v_r\Delta v_{\phi}\right>,
\end{equation}
\begin{equation}
\sigma_{\phi\phi}\equiv-\left<\sigmad\right>\left<\Delta v_{\phi}^2\right>-\left<\Delta\sigmad\Delta v_{\phi}^2\right>,
\end{equation}
represent the so called Reynolds stress. By using a closure relation $\left<\Delta v_r^2\right>=\left<\Delta v_{\phi}^2\right>=\cd^2$, we obtain the effective pressure gradient force as follows \citep{Shariff2011}:
\begin{equation}
\frac{1}{r}\frac{\partial\left(r\sigma_{rr}\right)}{\partial r}-\frac{\sigma_{\phi\phi}}{r}=-\frac{\partial\left(\cd^2\left<\sigmad\right>\right)}{\partial r}+\frac{1}{r}\frac{\partial\left(r\sigma_{rr}'\right)}{\partial r}-\frac{\sigma_{\phi\phi}'}{r},
\end{equation}
\begin{equation}
\sigma_{rr}'\equiv-\left<\Delta\sigmad\Delta v_r^2\right>,
\end{equation}
\begin{equation}
\sigma_{\phi\phi}'\equiv-\left<\Delta\sigmad\Delta v_{\phi}^2\right>.
\end{equation}
We neglect the terms $\sigma_{rr}',\sigma_{r\phi},\sigma_{\phi\phi}'$ for simplicity since a closure relation on these terms is uncertain. Moreover, we only consider the case that the dust grains are small and the friction is strong enough and assume $\Delta v_r=\Delta u_r$, $\Delta v_{\phi}=\Delta u_{\phi}$. Equations (\ref{eq:ap:momr_av}) and (\ref{eq:ap:momphi_av}) become
\begin{align}
\frac{\partial\left(\left<\sigmad\right>\left<v_r\right>\right)}{\partial t}&+\frac{1}{r}\frac{\partial}{\partial r}\left[r\left<\sigmad\right>\left(\left<v_r\right>-\frac{D}{\left<\sigmad\right>}\frac{\partial\left<\sigmad\right>}{\partial r}\right)\left<v_r\right>\right]\notag\\
=&\left<\sigmad\right>\frac{\left<v_{\phi}\right>^2}{r}-\frac{\partial\left(\cd^2\left<\sigmad\right>\right)}{\partial r}-\left<\sigmad\right>\frac{\partial}{\partial r}\left(\left<\Phi\right>-\frac{GM_{\ast}}{r}\right)-\left<\Delta\sigmad\frac{\partial\Delta\Phi}{\partial r}\right>-\left<\sigmad\right>\frac{\left<v_r\right>-\left<u_r\right>}{\tstop}\notag\\
&+\frac{\partial}{\partial t}\left(D\frac{\partial\left<\sigmad\right>}{\partial r}\right)+\frac{1}{r}\frac{\partial}{\partial r}\left(r\left<v_r\right>D\frac{\partial\left<\sigmad\right>}{\partial r}\right),\label{eq:ap:momr_av_b}
\end{align}
\begin{align}
\frac{\partial\left(\left<\sigmad\right>\left<v_{\phi}\right>\right)_t}{\partial t}&+\frac{1}{r}\frac{\partial}{\partial r}\left[r\left<\sigmad\right>\left(\left<v_r\right>-\frac{D}{\left<\sigmad\right>}\frac{\partial\left<\sigmad\right>}{\partial r}\right)\left<v_{\phi}\right>\right]\notag\\
&=-\frac{\left<\sigmad\right>\left<v_{\phi}\right>}{r}\left(\left<v_r\right>-\frac{D}{\left<\sigmad\right>}\frac{\partial\left<\sigmad\right>}{\partial r}\right)-\left<\sigmad\right>\frac{\left<v_{\phi}\right>-\left<u_{\phi}\right>}{\tstop}.\label{eq:ap:momphi_av_b}
\end{align}
The fourth term on the right hand side of Equation (\ref{eq:ap:momr_av_b}) stands for gravity from the density fluctuation generated by the turbulence. The self-gravity is determined by the volume integral of the density. Here, we assume that the volume-integrated density fluctuation is small and neglect the fourth term on the right hand side of Equation (\ref{eq:ap:momr_av_b}). In this work, we follow \citet{Cuzzi1993} and neglect the sixth term on the right hand side by assuming the term is smaller than the time derivative of $\left<\sigmad\right>\left<v_r\right>$. The seventh term represents the advection of the linear momentum $\left<\Delta \sigmad\Delta v_r\right>$ with the mean velocity $\left<v_r\right>$. This term is the same order of the advection of $\left<\sigmad\right>\left<v_r\right>$ along the diffusion flow that appears in the left hand side. As a consequence, we rearrange Equations (\ref{eq:ap:momr_av_b}) and (\ref{eq:ap:momphi_av_b}), and obtain the equations describing the time evolution of the mean velocity:
\begin{align}
\left<\sigmad\right>\left[\frac{\partial\left<v_r\right>}{\partial t}+\left(\left<v_r\right>-\frac{D}{\left<\sigmad\right>}\frac{\partial\left<\sigmad\right>}{\partial r}\right)\frac{\partial \left<v_r\right>}{\partial r}\right]=&\left<\sigmad\right>\frac{\left<v_{\phi}\right>^2}{r}-\frac{\partial\left(\cd^2\left<\sigmad\right>\right)}{\partial r}-\left<\sigmad\right>\frac{\partial}{\partial r}\left(\left<\Phi\right>-\frac{GM_{\ast}}{r}\right)\notag\\
&-\left<\sigmad\right>\frac{\left<v_r\right>-\left<u_r\right>}{\tstop}+\frac{1}{r}\frac{\partial}{\partial r}\left(r\left<v_r\right>D\frac{\partial\left<\sigmad\right>}{\partial r}\right),\label{eq:ap:eomr_av}
\end{align}
\begin{equation}
\left<\sigmad\right>\left[\frac{\partial\left<v_{\phi}\right>}{\partial t}+\left(\left<v_r\right>-\frac{D}{\left<\sigmad\right>}\frac{\partial\left<\sigmad\right>}{\partial r}\right)\frac{\partial \left<v_{\phi}\right>}{\partial r}\right]=-\frac{\left<\sigmad\right>\left<v_{\phi}\right>}{r}\left(\left<v_r\right>-\frac{D}{\left<\sigmad\right>}\frac{\partial\left<\sigmad\right>}{\partial r}\right)-\left<\sigmad\right>\frac{\left<v_{\phi}\right>-\left<u_{\phi}\right>}{\tstop}\label{eq:ap:eomphi_av}
\end{equation}
From Equation (\ref{eq:ap:eomphi_av}), we obtain
\begin{equation}
\left<\sigmad\right>\left[\frac{\partial\left( r\left<v_{\phi}\right>\right)}{\partial t}+\left(\left<v_r\right>-\frac{D}{\left<\sigmad\right>}\frac{\partial\left<\sigmad\right>}{\partial r}\right)\frac{\partial \left(r\left<v_{\phi}\right>\right)}{\partial r}\right]=-\left<\sigmad\right>r\frac{\left<v_{\phi}\right>-\left<u_{\phi}\right>}{\tstop},\label{eq:app_speangmon_av}
\end{equation}
which is equivalent to Equation (\ref{eq:speang_rphi}) since the mean specific angular momentum is $r\left<v_{\phi}\right>$. We can also obtain the evolutionary equation for the angular momentum of the dust that is equivalent to Equation (\ref{eq:angmonchange_sec2}) by using Equations (\ref{eq:app_diffusion}) and (\ref{eq:app_speangmon_av}):
\begin{equation}
\frac{\partial\left(\left<\sigmad\right>r \left<v_{\phi}\right>\right)}{\partial t}+\frac{1}{r}\frac{\partial}{\partial r}\left[r\left(\left<v_r\right>-\frac{D}{\left<\sigmad\right>}\frac{\partial\left<\sigmad\right>}{\partial r}\right)\left<\sigmad\right>r\left<v_{\phi}\right>\right]=-r\left<\sigmad\right>\frac{\left<v_{\phi}\right>-\left<u_{\phi}\right>}{\tstop}.
\end{equation}
If we use Equations (\ref{eq:eocgas}) and (\ref{eq:eomgas}) for gas, we can derive an equation equivalent to Equation (\ref{eq:new_angmon_change_w_vis}). Therefore, the equations formulated above hold the total angular momentum conservation. We omit the brackets representing the averaged value in the main part of this paper for convenience.

\section{Static mode in self-gravitating disks: No friction, Diffusion or Viscosity}\label{ap:staticmode}
In this appendix, we briefly summarize the dispersion relation of the dusty-gas disk for the case without the friction, the dust diffusion and turbulent gas viscosity. The dust and the gas interact with each other only through the self-gravity. In this case, the linearized equations of motion are the followings:
\begin{equation}
n\delta u_x=2\Omega\delta u_y-\frac{\cs^2}{\Sigma_0}ik\delta\Sigma-ik\delta\Phi, \label{eq:b20}
\end{equation}
\begin{equation}
n\delta u_y=-\frac{\Omega}{2}\delta u_x,\label{eq:b21}
\end{equation}
\begin{equation}
n\delta v_x=2\Omega\delta v_y-\frac{\cd^2}{\sigmadup}ik\delta\sigmad-ik\delta\Phi,\label{eq:b22}
\end{equation}
\begin{equation}
n\delta v_y=-\frac{\Omega}{2}\delta v_x.\label{eq:b23}
\end{equation}
From the above four equations and Equations (\ref{eq:lin-eocgas}), (\ref{eq:lin-eocdust}), (\ref{eq:lin-poisson}) with $D=0$,  we obtain the six-order equation for $n$:
\begin{equation}
n^2F_{\mathrm{DWs}}(n)=0, \label{eq:dispDW}
\end{equation}
\begin{equation}
F_{\mathrm{DWs}}(n,k)\equiv\left(n^2+\Omega^2+\cs^2k^2-2\pi G\Sigma_0k\right)\left(n^2+\Omega^2+\cd^2k^2-2\pi G\sigmadup k\right)-4\pi^2G^2\Sigma_0\sigmadup k^2
\end{equation}
Equation (\ref{eq:dispDW}) gives four density waves ($F_{\mathrm{DWs}}(n,k)=0$) and two static modes ($n=0$). In a case where we ignore the self-gravity of the gas, which corresponds to $\Sigma_0\to0$ in the above equation, $F_{\mathrm{DWs}}(n,k)=0$ gives
\begin{equation}
-n^2=\Omega^2+\cs^2k^2,
\end{equation}
\begin{equation}
-n^2=\Omega^2+\cd^2k^2-2\pi G\sigmadup k.
\end{equation}
The latter one is the dispersion relation of the density wave in the dust disk \citep[see also,][]{Youdin2011}. We can also reproduce the dispersion relation of the density wave in the gas disk from $F_{\mathrm{DWs}}(n,k)=0$ in a case where we ignore the self-gravity of the dust, which means that the dispersion relation $F_{\mathrm{DWs}}(n,k)=0$ describes two dust density waves and two gas density waves. The density wave becomes unstable if the self-gravity is strong enough. This unstable density wave is the classical gravitational instability.

A static mode is a steady solution of the linearized equations. This mode is referred to as a ``neutral mode" in \citet{Youdin2011}. For both static modes we obtain here, $\delta u_x=\delta v_x=0$ (Equations (\ref{eq:b21}) and (\ref{eq:b23})) and the radial force balance holds (Equations (\ref{eq:b20}) and (\ref{eq:b22})). For one static mode, the gas and the dust have different azimuthal velocity while they have the same azimuthal velocity for the other static mode. The former static mode is destabilized by the friction. The friction does not change the latter static mode since there is no relative motion between the gas and the dust. The latter mode is destabilized once we consider both friction and viscosity.

\section{Linear analysis in a thin dust disk}\label{ap:dustsubdisk}
In Section \ref{sec:analysis_dustsubdisk}, we discuss the stability in the thin dust disk by assuming the vertical density profiles. However, the range of the vertical integration to obtain the surface densities, $\Sigma_0'$ and $\sigmadup'$, is uncertain. Thus, we here show results for cases where the range of the vertical integration is $-\hd\leq z \leq \hd$ and $-2\hd\leq z \leq 2\hd$.

\begin{figure}[h]
	\begin{center}
		\includegraphics[width=0.5\columnwidth]{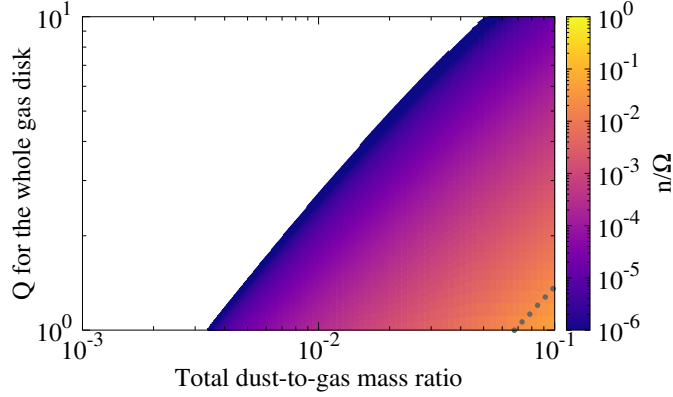}
	\end{center}
    		\caption{Maximum growth rate of the instabilities for $\taus=0.1$ and $\alpha=10^{-4}$. The range of the vertical integration is $-\hd\leq z \leq \hd$. The horizontal axis is the total dust-to-gas mass ratio $\sigmadup/\Sigma_0$. The vertical axis is Toomre's $Q$ value for the whole gas disk, that is, $Q=\cs\Omega/\pi G\Sigma_0$. The color represents the maximum growth rate normalized by the angular velocity $\Omega$. TVGI is the most unstable mode in the colored region above the gray dotted line, while the dust GI mode grows fastest below the gray dotted line.}
   		 \label{fig:taus01alph1e-4-growthrate_w_viscosity_w_thickness_1hd}
\end{figure}

Figure \ref{fig:taus01alph1e-4-growthrate_w_viscosity_w_thickness_1hd} shows the maximum growth rate as a function of total dust-to-gas mass ratio $\sigmadup/\Sigma_0$ and $Q=\cs\Omega/\pi G\Sigma_0$ for the case with the integral in $-\hd\leq z \leq \hd$. The normalized stopping time $\taus$ and the strength of the turbulence $\alpha$ are set to be $0.1$ and $10^{-4}$, which is same with those in Figure \ref{fig:taus01alph1e-4-growthrate_w_viscosity_w_thickness}. In this case, $\tilde{Q}/Q$ is about 42, and $\sigmadup'/\Sigma_0'$ is about 28 times larger than the total dust-to-gas mass ratio $\sigmadup/\Sigma_0$. TVGI grows fastest above the gray dotted line. The maximum value of $Q$ for each total dust-to-gas mass ratio is smaller than that seen in Figure \ref{fig:taus01alph1e-4-growthrate_w_viscosity_w_thickness}, which is because $\tilde{Q}/Q$ is larger, and the gas disk is self-gravitationally more stable. For the same reason, the region where the secular GI is fastest growing mode does not appear, and the dust GI grows fastest below the gray dotted line. 
\begin{figure}[h]
	\begin{center}
		\includegraphics[width=0.5\columnwidth]{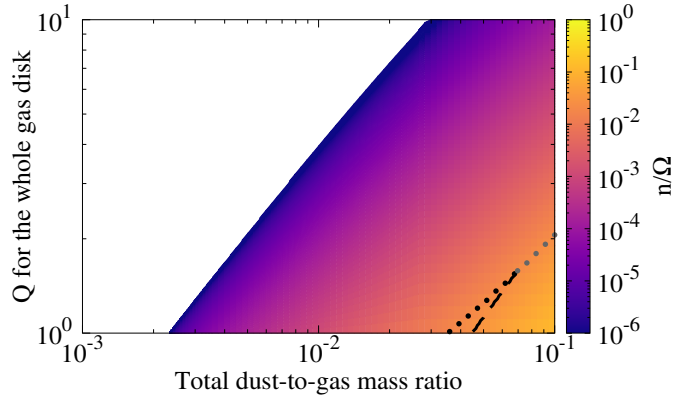}
	\end{center}
    		\caption{Maximum growth rate of the instabilities for $\taus=0.1$ and $\alpha=10^{-4}$. The range of the vertical integration is $-2\hd\leq z \leq 2\hd$. The horizontal axis, the vertical axis and the color are same in Figure \ref{fig:taus01alph1e-4-growthrate_w_viscosity_w_thickness_1hd}. In the colored region above the black and gray dotted lines, TVGI is the most unstable mode. The secular GI is the fastest growing mode in the region enclosed by the black dotted and dashed lines. The dust GI mode becomes the most unstable mode below the gray dotted and dashed line.}
   		 \label{fig:taus01alph1e-4-growthrate_w_viscosity_w_thickness_2hd}
\end{figure}

Figure \ref{fig:taus01alph1e-4-growthrate_w_viscosity_w_thickness_2hd} is a result for the case with the integral in $-2\hd\leq z \leq 2\hd$. The ratio $\tilde{Q}/Q$ is about 21, and $\sigmadup'/\Sigma_0'$ is about 20 times larger than $\sigmadup/\Sigma_0$ in this case. The secular GI is the fastest growing mode in the region enclosed by the black dotted and dashed lines. TVGI grows fastest  in most of the colored region, which is qualitatively same with Figures \ref{fig:taus01alph1e-4-growthrate_w_viscosity_w_thickness} and \ref{fig:taus01alph1e-4-growthrate_w_viscosity_w_thickness_1hd}.




\bibliographystyle{aasjournal}
\bibliography{rttominaga2019}

%
%
%


\end{document}